\documentclass[]{interact}

\usepackage{epstopdf}
\usepackage[caption=false]{subfig}

\usepackage[longnamesfirst,sort]{natbib}
\bibpunct[, ]{(}{)}{;}{a}{,}{,}
\renewcommand\bibfont{\fontsize{10}{12}\selectfont}

\theoremstyle{plain}

\theoremstyle{definition}

\theoremstyle{remark}

\usepackage{hyperref}
\usepackage{longtable}
\usepackage{pdflscape}
\usepackage{lmodern}

\usepackage[longnamesfirst,sort]{natbib}
\begin{document}


\title{AMACA: Astronomy education with a Multi-sensory, Accessible, and Circular Approach}

\author{
\name{Rachele Toniolo\textsuperscript{a,b,c}*\thanks{rachele.toniolo@inaf.it}, Anita Zanella\textsuperscript{e,d,c}, Andrea Cottinelli\textsuperscript{e,f}, Giovanni Liuzzi\textsuperscript{e}, Sara Ricciardi\textsuperscript{d,c}, Massimo Grassi\textsuperscript{g}, Stefano Delle Monache\textsuperscript{h}}
\affil{\textsuperscript{a}Dipartimento di Fisica e Astronomia, Università di Bologna, Via Gobetti 93/2, 40129, Bologna, Italy;
\\\textsuperscript{b}INAF - Istituto di Radioastronomia, Via Gobetti, 101, 40129, Bologna, Italy;
\\\textsuperscript{c}IAU Office of Astronomy for Education Center Italy;
\\\textsuperscript{d}INAF - Osservatorio di Astrofisica e Scienza dello Spazio di Bologna, Via Gobetti 93/3, 40129, Bologna, Italy; 
\\\textsuperscript{e}INAF - Osservatorio Astronomico di Padova, Vicolo dell'Osservatorio 5, I-35122 Padova, Italy; 
\\\textsuperscript{f}IAU OAE National Astronomy Education Coordinator Team for Italy; 
\\\textsuperscript{g}Department of General Psychology, University of Padua, Via Venezia 8, 35131, Padua, Italy; 
\\\textsuperscript{h}IRCAM STSM Lab, Sound Perception and Design group, 1 Place Igor Stravinsky, 75004 Paris, France}
}

\maketitle

\begin{abstract}
The AMACA project (``Astronomy education with a Multi-sensory, Accessible, and Circular Approach'') develops multi-sensory activities for accessible education and public engagement in the field of Astronomy.  Despite promising innovations, existing resources are often poorly documented, designed for one-time events, or too expensive to scale. Additionally, most initiatives lack interdisciplinary collaboration and user testing, and dissemination is often restricted to their original creators. AMACA seeks to overcome these challenges by creating multi-sensory activities for education and public engagement. 
A circular approach informs the educational structure of AMACA: (1) a PhD course on the multi-sensory design of Astronomy outreach delivers several educational activities in the form of hands-on workshops, with the support of astronomers, psychologists, and various organizations for the visually impaired and the deaf; (2) PhD candidates train High School (HS) students in delivering the educational activities; (3) HS students lead these hands-on workshops at the public Astronomy Festival ``The Universe in All Senses''; (4) teachers of primary and secondary education cycles (i.e. 6 - 19 years old) are trained by the HS students and run the educational activities in their classes.
Also, AMACA develops and tailors tools to guide the project development, to collect and track the participants' learning process.
Key findings include: PhD candidates improved their ability to communicate astronomy to non-experts (such as the general public and school pupils) and developed greater awareness of accessibility in the field; HS students gained a more emotionally engaging understanding of astronomy; and the general public learned about research findings and methods while also recognizing increased accessibility for visually impaired individuals. Teachers expressed satisfaction with the flipped roles approach and hands-on learning.
Overall, AMACA's multi-sensory, hands-on approach enhances the accessibility and engagement of astronomy education for PhD candidates, HS students, the general public, and teachers alike.
\end{abstract}


\section{Introduction}

Astronomy is commonly perceived as a visual science. In a centuries-old stereotypical portrayal, astronomers are depicted as individuals gazing at the stars. Although this does not happen anymore — at least for professional astronomers — astronomy is still communicated mostly through spectacular images and videos. The numerical data collected by telescopes on the ground (e.g., the Very Large Telescope\footnote{https://www.eso.org/public/teles-instr/paranal-observatory/vlt/} in the Atacama Desert) and in space (e.g., the recently launched James Webb Space Telescope\footnote{https://webbtelescope.org/home}) are turned into inspiring images, while computer simulations represent numerical data as animations of astrophysical phenomena. Such images and videos also include the visualization of data that are ``invisible'', such as electromagnetic radiation outside the visible domain (e.g., ultraviolet, infrared, radio wavelengths) and phenomena that do not emit light (e.g., dark matter). Despite the fascination that these representations produce for the public, this visually-oriented approach naturally excludes audiences who are blind or visually impaired (BVI) and, more generally, those who may prefer non-visual communication.

On the educational side, astronomy has several crucial characteristics that make it powerful for communicating in formal and informal educational contexts.
First, it is perceived as a fascinating science by most of the public, due to the wonder that distant objects in the Universe inspire in everyone. This facilitates to engage a significant fraction of the public and/or students who are not interested in science initially. Second, it is considered a gateway science for its multi-disciplinary nature \cite{salimpour2021gateway}. In fact, astronomy education includes various Science, Technology, Engineering, Mathematics (STEM) disciplines, such as math, physics, geology, and biology, but also philosophy and art (STEAM instead of STEM, where the ``A'' stands for Art). This provides an ideal playground for designing multi-sensory and multi-disciplinary resources. Finally, astronomy is perceived as the visual science par excellence, making it a good challenge and starting point for a paradigm shift to make science multi-sensory and accessible to everyone.

In recent years, innovative approaches have been developed to make astronomy communication multi-sensory \cite{Perez-Montero2019, Varano2023, Noel-Storr2022, Foran2022, Harrison2022, Varano2024}. An increasing number of projects focus on developing tactile resources \cite{Bonne2018, Paredes-Sabando2021, Arcand2022} and/or audible resources \cite{Quinton2016, Tomlinson2017, Bieryla2020, Elmquist2021, Garcia-Benito2022, Bardelli2022, Harrison2023, Guiotto2024}. In particular, the process of turning data into sound, called ``sonification'', is gaining momentum in the astronomical community with applications in both astronomy research and communication (see \cite{Zanella2022} for a review and the Data Sonification Archive\footnote{Data Sonification Archive: \href{https://sonification.design/}{https://sonification.design/}} for a repository of astronomical data sonification applications, \cite{Lenzi2020}). These pioneering efforts have shown that a multi-sensory approach makes astronomy communication not only more inclusive but also more engaging for everyone.

While encouraging, these attempts to make astronomy communication multi-sensory still face limitations preventing them from becoming widely shareable. First, many resources are being developed for specific one-time events \cite{Bieryla2020, Elmquist2021}; they often lack substantial documentation and are not easily accessible. Second, several of these resources, particularly those adopting tactile approaches, require materials that are difficult to manufacture, expensive, and/or time-consuming to produce for large audiences \cite{Bonne2018, Usuda-Sato2019, Paredes-Sabando2021, Perez-Montero2022, Arcand2022}. Third, many resources have been created by teams of astronomers alone, without collaboration from multi-disciplinary experts (e.g., sound designers, psychologists) and have not been properly tested with users before being adopted by the public. Lastly, such resources are generally disseminated by the same professionals who created them, making them difficult to share widely and delivered by non-specialists \cite{Harrison2023}.

To overcome these limitations, we created AMACA\footnote{AMACA: \href{https://sites.google.com/inaf.it/amaca/home-page}{https://sites.google.com/inaf.it/amaca/home-page}}, which stands for ``Astronomy education with a Multi-sensory, Accessible, and Circular Approach''. AMACA is a project aiming to raise awareness about the potential of a multi-sensory approach to Astronomy through the development of multi-sensory activities\textbf{\footnote{see the AMACA project website for the description of activities: https://www.universointuttiisensi.it/workshops/} }(i.e., that use two or more senses to deliver the scientific content) for public engagement and astronomy education. Such activities are designed in the course of a structured educational path involving professional astronomers, psychologists, sound designers, HS science teachers, astronomy educators, sensory-disability experts, PhD candidates, high-school (HS) students, and teachers of all school cycles. All the resources are created using affordable, ready-made and possibly re-usable materials, they are explored with and evaluated by the participants, and they are properly documented to be widely disseminated by non-specialists (e.g., teachers, science communicators). The resources developed in the context of AMACA are meant to be used not only at science festivals and public engagement events, but also in classrooms during science lectures. To do so, every activity is widely described and available in the free repository PlayINAF\footnote{PlayInaf: \href{https://play.inaf.it/tag/universo-in-tutti-i-sensi/}{https://play.inaf.it/tag/universo-in-tutti-i-sensi/}}.
This approach refers to the Universal Design for Learning (UDL) methodology, an educational model that promotes the design of learning experiences from the outset to accommodate diversity, rather than retrofitting solutions for specific needs.

\section{AMACA: an educational path}
\label{subsec:chain}

The AMACA project follows an educational path aimed at sharing scientific culture in an atmosphere of cooperation and inclusion. Through this approach, AMACA transformed the traditional dynamic between science communicators and the target audience. Those who were initially considered the audience became active participants, contributing directly to the design and execution of hands-on workshops. This shift ensures their sustainability and fosters inclusion, equity, and adaptability.

Two important key features have been integrated into the process. First,  a bottom-down educational approach has been used instead of a top-down one. All the phases listed next are characterized by the fostering of free experimentation to promote autonomy, one of the main components to promote meaningful and effective learning \cite{deci2000and}.

Secondly, in several activities developed in the context of AMACA we focus on methodologies and practices (e.g., how exoplanet transits are discovered with the light curve method; how the chemical composition of comets and planets’ atmosphere are investigated, ...). Conveying to the public how research is done rather than communicating only spectacular results shortens the distance gap between the scientific research community and society \cite{potochnik2025public}, the primary goal of public engagement with science \cite{weingart2021public}.

AMACA started in 2021 and it was initially envisioned as an Astronomy Festival where hands-on workshops were led by trained HS students. They proved to be exceptional science communicators and effective in successfully engaging their primary audience, families. 
In 2022, astrophysics PhD candidates were involved in the project. As young scientists, they are typically well-versed in astrophysics but often need more professional experience in science communication and education, in particular with non-experts. Therefore, a university course was established to provide them with the tools and methodologies to design effective public engagement activities.
Teachers were directly involved in 2023, as it became evident that the activities created in the context of AMACA could serve as valuable resources for classroom activities. Therefore a dedicated training course aimed at engaging and encouraging teachers to integrate multi-sensory and hands-on workshops into their teaching practices became part of the AMACA educational path.

So far the AMACA project has run four editions of the Astronomy Festival which were attended by more than 10,000 visitors, involved 220 HS students (from 2021 to 2024), 45 astrophysics PhD candidates in three courses (2022-2024), and 18 teachers of all educational cycles (2023-2024).

\begin{figure}
    \centering
    \includegraphics[width=\textwidth]{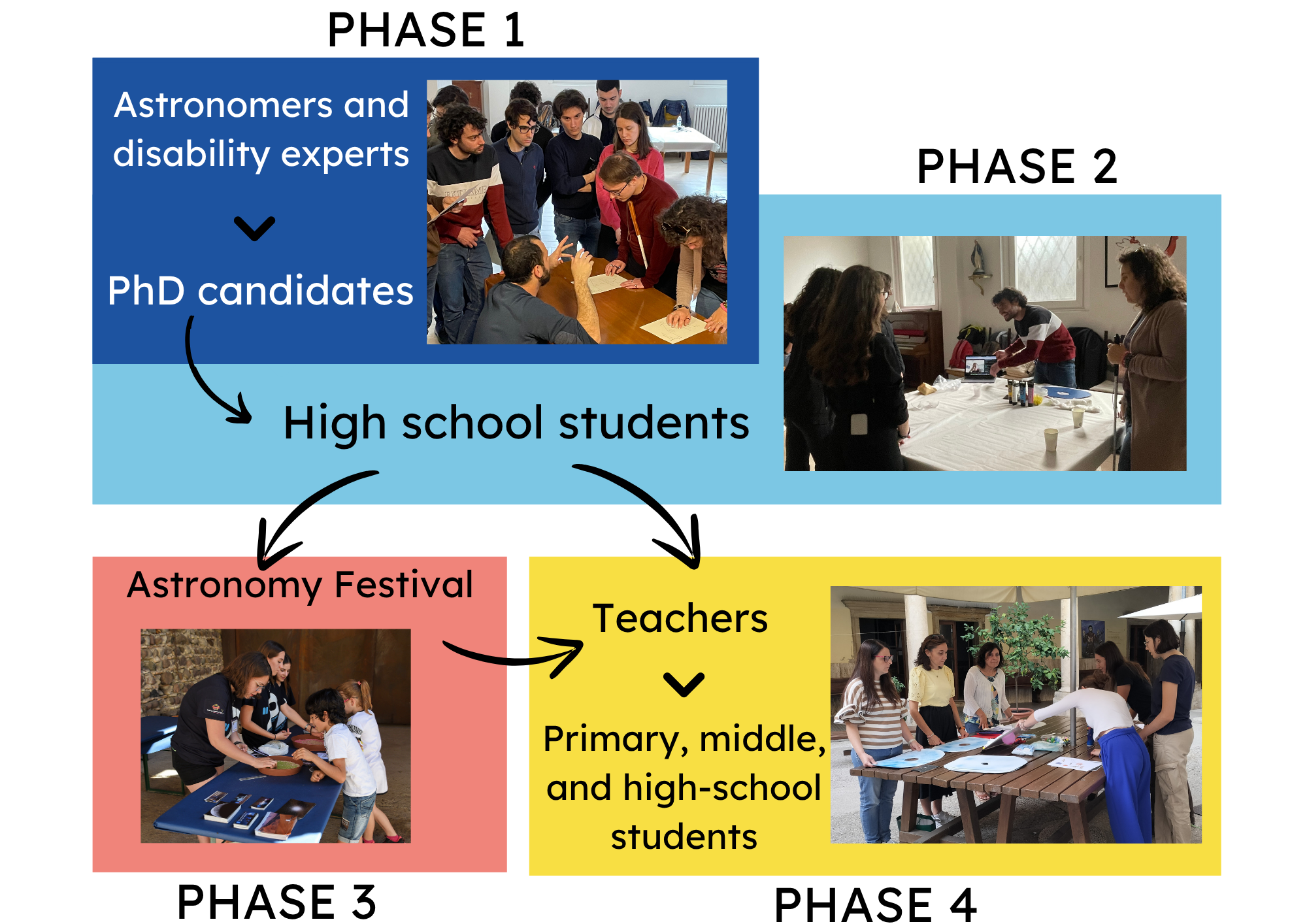}
    \caption{An educational path to share scientific culture in an atmosphere of cooperation and inclusion. Summary of the phases that are at the basis of the AMACA project.} 
    \label{fig:amaca}
\end{figure}

\begin{figure}
    \centering
    \includegraphics[width=\textwidth]{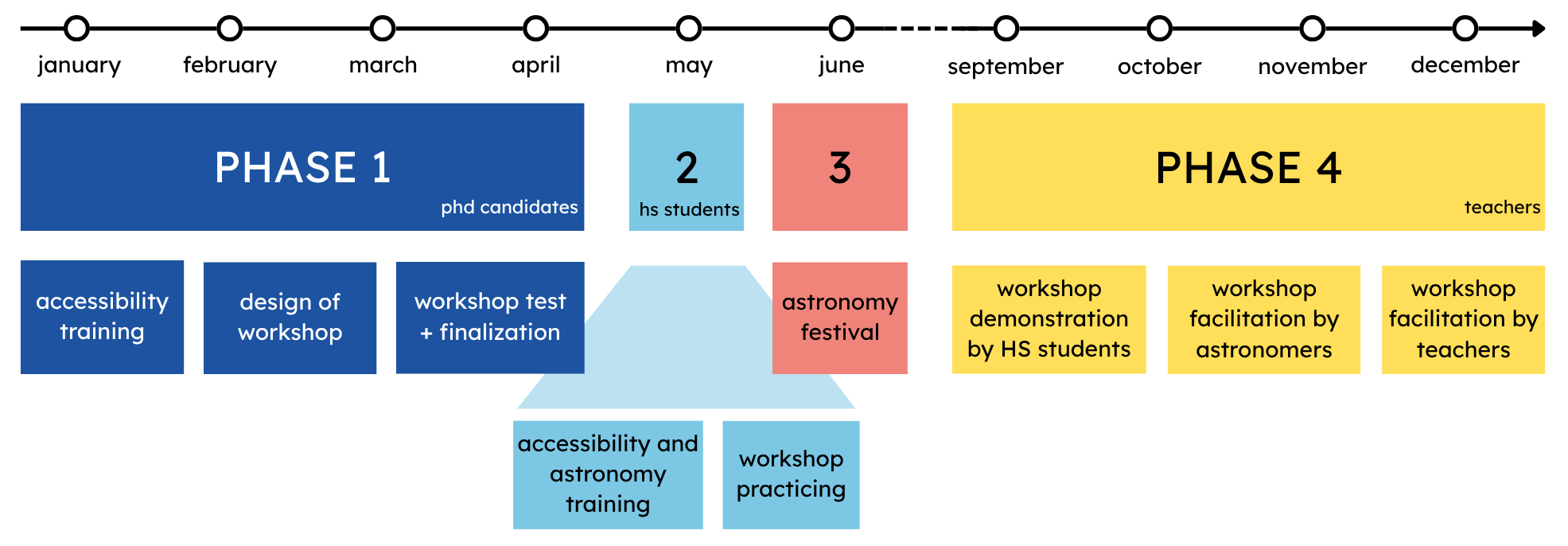}
    \caption{The timeline of the AMACA educational path.} 
    \label{fig:timeline}
\end{figure}

The AMACA educational path consists of four phases, which are visually outlined in Figures \ref{fig:amaca}, \ref{fig:timeline} and detailed below. For each section, we stress the accessibility strategy introducing a brief paragraph that describes the actions adopted.

\subsection*{Phase 1 - From astronomers and disability experts to PhD students}
\textbf{\\Lecturers: astronomers with PhD, psychologists, astronomy educators, disability experts, science teachers at HS\\Target: PhD candidates from the Physics and Astronomy Department of the Italian Universities of Padova and Bologna}\\

A university course is held for Astrophysics PhD candidates, focusing on developing interactive and multi-sensory outreach activities. The class is part of the regular curriculum of the Physics and Astronomy department. The lecturers are astronomers who work in the outreach and science education field at the National Institute for Astrophysics (INAF), and as science teachers at high-school. Some of the lectures are held in collaboration with psychologists who work in the area of educational psychology and disabilities, and representatives of the Italian Union of the Blind and Visually Impaired (UICI - Unione Italiana Ciechi e Ipovedenti). The PhD candidates are from the Physics and Astronomy Department of the Italian Universities of Padova and Bologna. The number of attendees depends on the year and ranges between 10 and 30.

The course has a duration of 18 hours and includes a final exam. It runs from February to May, with lectures of two hours each, every two weeks.
During the course, the students are guided to develop hands-on activities related to astrophysics that involve at least two senses. To guarantee accessibility, the activities are original and conceived from the start with this objective in mind, i.e., PhD candidates do not take already existing activities and modify them to add accessibility. Each year, the activities revolve around a different theme. From 2022 to 2024, the themes have been gravity (``The attractive Universe''), multisensoriality (``The Universe in all senses''), and time (``On the trail of time'').

The course accommodates two cycles of evaluation and design. 1) Once the activities are designed, they are tested with primary school students and people with sensory disabilities, mainly individuals with BVI; 2) the students improve the workshops according to the participants' feedback, and properly document the activities. At the end of the course, the PhD candidates provide HS students with a training on the designed educational activities, with a careful focus on both the scientific content and its presentation mode with the public  (see ``Phase 2''). The activities are facilitated at the Astronomy Festival ``The Universe in all senses'' (yearly in June, see ``Phase 3'') and delivered to school teachers in Fall (see ``Phase 4''). The outline of the course is reported in Table \ref{tab:course}.

``The many looks of a galaxy'' is one good example of a hands-on activity authored through the AMACA process (tutorial available at this \href{https://play.inaf.it/en/how-many-looks-for-a-galaxy/}{link}). It was designed during the 2023 course by four PhD candidates from the University of Bologna. The activity aims to convey the importance of observing celestial objects at different wavelengths to obtain a comprehensive understanding of their physics. Participants are asked to use pieces of fabric with various textures to reproduce different kinds of electromagnetic radiation emitted by the galaxy Centaurus A. Each wavelength represented by a texture fabric is also associated with one musical instrument playing a melody, leading to the idea that only when the instruments play together, you can recognized the song that is played and enjoy the music. The activity was tested with three primary-school pupils of 8 years old, it was learnt by 4 HS students, and presented at the third edition (2023) of the Astronomy Festival to a few thousand visitors.

\subsubsection*{\textbf{Accessibility strategy.}} 
The course was collaboratively designed and co-taught with accessibility experts. In 2024 the Deputy Rector for Inclusion and Disabilities, the Head of the Learning Lab for Deaf Children at the University of Padova, and the President of the UICI in Padova were involved. Awareness of accessibility issues is taught by employing experiential and playful learning activities to provide a first-person understanding of sensory limitations, by systematically occluding different sensory channels. Moreover, two lectures were specifically dedicated to accessibility and techniques for making laboratories accessible.
For many of the PhD candidates, this was their first direct encounter with sensory disability situations.

\subsection*{Phase 2 - From PhD candidates to HS students}
\textbf{\\Lecturers: PhD candidates and lecturers of Phase 1\\Target: HS students from a variety of Italian schools, including classical, scientific and linguistic lyceums, and technical institutes}\\

In this phase, the PhD candidates (assisted by the lecturers of the course described in ``Phase 1'') become trainers and teach HS students the activities they created and how to approach the public. The training is part of the ``Pathways for Transversal Skills and Orientation'' (PCTO), a curricular program required by Italian high schools that aims to provide students with a more comprehensive educational experience that goes beyond classroom learning, to bridge the gap between academic knowledge and real-world application by focusing on transversal skills and orientation. Institutions, associations, and companies can offer PCTO programs through the stipulation of an agreement with each specific school. Students can then enroll in the activities that they prefer and gain a number of credits that depends on the number of hours required to complete the PCTO program. The PCTO program proposed by AMACA lasts for 30 hours. 

The AMACA training PCTO program is split in two parts. In the first part over a compressed two-day lecture (8 hours), HS students receive an introduction on the basic astronomical concepts underlying the educational activities to learn. UICI and ENS (National Deaf Organization, Ente Nazionale Sordi) representatives also introduce best practices for an effective and respectful communication with people with sensory disabilities. HS students are also trained on how to welcome and engage the public with the activities.

During the second part of the training, HS students are trained on one of the activities developed by PhD candidates. HS students, divided into small groups, pick the activity that best matches their interests and attitudes. The overall approach exploits a hands-on approach, in which HS students first experience the activity as public and then go deeper into the astronomy content with the PhD candidates. The aim is to give HS students the materials and resources to rehearse the activity on their own. During this phase, HS students practice the activity chosen multiple times with experts until they are proficient. A summary of the training is reported in Table \ref{tab:training_highschool} (lecture 1).

Depending on the year, participation to the AMACA PCTO program varies between 40 to 60 HS students, coming from 10 high schools of various educational backgrounds located in three main cities in the North of Italy (Verona, Mantova, Brescia, and their respective provinces).
 
\subsubsection*{\textbf{Accessibility strategy.}}
As in ``Phase 1'', students were involved in experiential and playful learning activities. The aim was to have them personally explore the characteristics and difficulties associated with sensory disabilities to develop accessible strategies for communication and dissemination.

In 2024, the ENS President of the Regional Council of Veneto, and his Italian Sign Language interpreter were involved. They guided the students in discovering hearing disabilities through a workshop aimed at stimulating questions and curiosity by sharing their own experiences.
The students then participated in empathy exercises that simulated the various degrees of visual disabilities, such as trying to read a book or navigate a space while wearing modified glasses that mimicked different levels of vision loss.

The phase of training about the workshops, operated by PhD candidates to HS students, was supervised by experts to ensure the proactive and successful communication in the interaction with sensory disabilities individuals.

\subsection*{Phase 3 - From HS students to the general public}
\textbf{\\Facilitators: HS students\\
Target: visitors of the Astronomy Festival "The Universe in all senses"}\\

The third phase happens in the context of the Astronomy Festival ``The Universe in all Senses'', which takes place each year in June in Castellaro Lagusello, a small village in the North of Italy. During the three-day event, the public can attend various activities, including the hands-on educational workshops developed in previous phases, outreach conferences, sky observations, astronomy-related shows, poetry performances, exhibitions, and wall-mapping projections. 
HS students are in charge of both leading the hands-on activities and welcoming the public at the Festival.

Since its first edition in 2021, the Festival hosted more than 10,000 visitors. All the activities are completely free for the visitors. The Festival's annual budget is approximately 30,000 euros, sourced from the sponsorships of private companies ($\sim 40\%$), financial support from public institutions and public funding calls ($\sim 55\%$), donations from individuals, and the sale of AMACA merchandise ($\sim 5\%$). The expenses are allocated as follows: $\sim 55\%$ for materials and technical services (e.g., materials for hands-on workshops, audiovisual technical support, design, printing and distribution of promotional materials, Italian Sign Language (LIS) interpreters, ambulance services, ...); $\sim 20\%$ for the cachet of the artists, speakers, and professionals invited to the Festival; $\sim 20\%$ for hospitality and travel expenses; $\sim 5\%$ for taxes and rentals.

\subsubsection*{\textbf{Accessibility strategy.}}
Accessibility has been an integral part of every stage of the Festival's design:
\begin{itemize}
    \item \emph{Contents:} Every activity included various sensory elements aiming not only to support accessibility but also to enrich the user's enjoyment and engagement. Exhibitions and screenings were accessible thanks to an integrated system of visual and sound cues. For conferences, workshops, and guided tours of the exhibitions, specific events were planned with simultaneous translation into Italian Sign Language.
    \item \emph{Navigation:} The Festival's structure was designed so that all locations were physically accessible and without architectural barriers. In addition, NaviLens\footnote{NaviLens: \href{https://www.navilens.com/en/}{https://www.navilens.com/en/}} codes were installed, scannable via app, to signal each location of the Festival, allowing participants to explore each location independently.
    \item \emph{Communication:} The Festival website, graphics, and social media channels followed the Web Content Accessibility Guidelines (WCAG)\footnote{The guidelines we followed are available at this link: \href{https://www.w3.org/TR/WCAG21/}{https://www.w3.org/TR/WCAG21/}}.
\end{itemize}

\subsection*{Phase 4 - From HS students to teachers}
\textbf{\\Lecturers: HS students and lecturers of Phase 1\\Target: teachers with various educational backgrounds.}\\

The last phase of the educational path is a course for teachers of different school levels\footnote{It is worth noting that in Italy there are no special schools specifically dedicated to the education of individuals with sensory disabilities. Instead, they attend mainstream schools, where they receive support from special education teachers who complement standard teaching with personalized interventions.} - from primary to high school - and disciplines, such as humanities, STEM, and special education. Their participation in the course is voluntary, yet recognized as part of the annual mandatory training. 

During the lessons, HS students and lecturers of Phase 1 who supervised the whole path, present the hands-on activities led during the Festival. Each teacher is trained on three different activities, which are chosen based on the age of the participants they address. The multi-disciplinarity and inclusiveness of the activities guarantee the participation of not only STEM teachers but also humanities and special education teachers.

The course is divided in three sessions.
1) HS students demonstrate how the hands-on activities are conducted and convey the challenges they faced with the public during the Festival, and how they solved them. This session takes place during one afternoon at the end of September or the beginning of October, depending on the availability of the participants.
2) Between October and December, one of the INAF astronomers holds one of the activities chosen by the teachers in their class. In this way, the teachers can assess how the students are engaged and how they react to such hands-on activity.
3) Teachers run one activity of their choice, while one INAF astronomer is present for support and to give feedback at the end of the session.

After every session is held with each teacher attending the course, there is one last online meeting with a final questions-and-answers session, where the teachers can share doubts, concerns, report challenges, ask for help, and give feedback. The total duration of the course is 10 hours. A summary of the course is reported in Table \ref{tab:course_teachers}.

The course aims to promote the use of hands-on multi-sensory activities in schools, fostering the experimentation of the approach both in teachers and their students. 

This phase was introduced for the first time in 2023. Seven teachers (two from primary school and five from middle school) attended the course that year, while eleven teachers (three from kindergarten, two from primary school, five from middle school, and one from high school) attended the course in 2024. The teachers mainly come from Verona, Mantova, and their provinces.

\subsubsection*{\textbf{Accessibility strategy.}} 
At this stage, teachers received brief training on the sonification of astronomical data, multi-sensory teaching, and resources that can be used for teaching activities in astronomy (AstroEDU\footnote{AstroEdu: \href{https://astroedu.iau.org/en/}{https://astroedu.iau.org/en/}} and PlayINAF\footnote{PlayINAF: \href{https://play.inaf.it/}{https://play.inaf.it/}}). It is assumed that they already had training in teaching and accessibility during their professional education. The focus is on applying their teaching skills in astronomy.

\section{Materials and methods}
\label{sec:methods}

We adopted different tools in each phase of the educational path, either to foster the development of the activities or to assess the impact of our approach. The tools are summarized in Table \ref{tab:tools_workshop} and described in detail below. Templates can be downloaded from the OSF Repository (\href{https://osf.io/7q69z/}{https://osf.io/7q69z/}).
\begin{table}
    \tbl{List of adopted tools.}
    {\begin{tabular}{p{3.7cm}p{0.8cm}p{7.5cm}} \hline 
         Tool & Phase & Aim\\ 
         \hline 
         AMACA Design Canvas&  1&  Support PhD candidates in the development of the activities\\ 
         Feedback Wheel&  1&  Collect feedback from participants testing the activities\\ 
         ICEBAGS&  1,2&  Create a list of good practices for leading hands-on activities\\ 
 Word-clouds& 2,3& Assessing the change of perspective of high-school students before and after the project\\
 Questionnaires& 3& Assessing the impact of using a multi-sensory approach for astronomy outreach and education\\
 Focus Group& 4& Assessing the perception of teachers about the hands-on approach of the activities and the role flipping\\
\bottomrule
\end{tabular}}
    \label{tab:tools_workshop}
\end{table}

\subsection{Materials used for the development of the activities}
\label{subsec:tools}

\subsubsection{AMACA Design Canvas}

The AMACA Design Canvas (Fig.~\ref{fig:temp}) is a conceptual design tool inspired by the Business Model Canvas \cite{osterwalder2010business} and the Sonification Design Canvas \cite{lenzi_canvas}, which we have adapted to suit our specific goals. The original Data Sonification Canvas was developed to support creators in designing new sonifications and in effectively communicating early design ideas to others \cite{lenzi2024designing}. It was introduced to the astronomy community during the Audible Universe workshops \cite{Harrison2022, Zanella2022, Misdariis2022}. 
The Sonification Design Canvas is structured around four key building blocks that represent essential areas in the sonification design process: ``use case'', ``mapping choices'', ``sonification approach'', and ``listening experience''. Each block is further divided into more detailed sub-fields. For instance, the ``use case'' block includes the sub-fields: ``users'', ``goals'', and ``context'' \cite{lenzi_canvas, lenzi2024designing}.
In developing the AMACA Design Canvas, we retained the overall structure and visual layout of the original Sonification Design Canvas but modified the titles and content of the blocks to better reflect our specific design needs. Each block is defined as follows:
\begin{itemize}
\item \textbf{Block 1: scientific aspects}. This is divided into three sub-fields: \textbf{content} (what scientific topic do you want to focus on? What scientific result/methodology/aspect do you find interesting to communicate, and what might be interesting or meaningful for your audience to receive?); \textbf{message} (what message do you want to convey? What would you like people to learn, understand, remember?); and \textbf{goals} (what do you want to achieve? Why should the audience come and care?) 

\item \textbf{Block 2: context and target}. This is divided into five sub-fields: \textbf{context} (where will the activity take place? In a classroom, science festival, research conference, planetarium?); \textbf{location} (will the activity be inside or outside? What kind of space is needed?); \textbf{duration} (how long is the activity meant to last?); \textbf{target audience} (who will be the main involved audience? Adults, children? With disability? Teachers, researchers, general public?); and \textbf{group/individual} (is the activity suited for a group of participants or for individual participation? What is the minimum and the maximum number of participants?). 

\item \textbf{Block 3: experience}. This is divided into three sub-fields: \textbf{approach} (what approach will you take? A ‘lecture’  approach or an ‘exploratory’ approach?); \textbf{technique} (what technique(s) will you use? Narration, hands-on, tinkering, exhibition? Analogic or digital? Will you use videos, projections?); and \textbf{material} (what material will you need? Projector, screen, speakers? Table, glue, paper? Copper tape, LEDs?). 

\item \textbf{Block 4: involved senses and notes}. This is divided into two sub-fields: \textbf{involved senses} (what senses will be involved? Sight and hearing? Touch, sight, and hearing? Smelling?); and a space for \textbf{other notes} (are there other important things to consider? E.g., is an Italian Language Sign interpreter needed? Could there be criticalities? Are there problems to solve or doubts? What is the expected cost for the realization of the activity?).
\end{itemize}

In Phase 1, PhD candidates use the AMACA Design Canvas as a supporting tool to conceptualize and design the activity. They complete the entire Canvas at the very beginning of the design process, to reflect on the main aspects of the workshop they want to create and plan it.
After the activities have gone through a first assembly phase, they are tested with students and individuals with sensory disabilities. This sample is chosen based on the personal contacts that the authors have with schools and associations in the region. The participation is voluntary.

After the test, PhD candidates are asked to adapt the AMACA Design Canvas, taking into account the changes implemented during the design process and the feedback received by the testers. This has the twofold goal of guiding the students toward the final design of their activity while, at the same time, allowing them to reflect on the process they went through, by comparing the first and second AMACA Design Canvas that they have created.

Once the activity is finalized, the students are asked to fill in a descriptive document that presents in detail the activity. The template of the document includes the following sections: title of the activity, brief description of the activity (max 10 lines), needed material (with a picture), activity preparation (group size and any preparation that needs to be done before the start), description of the activity development (with pictures if available and/or needed), description of the physical processes (a final section meant to give a more comprehensive explanation of the scientific aspects at the basis of the activity and useful links for a more in-depth study). The document is used both by HS students to learn and deliver the workshops during the Festival and by the teachers during the training. It is also uploaded on the INAF repository PlayINAF (\href{play.inaf.it}{play.inaf.it}) to allow anyone to recreate the activity.\footnote{\textbf{See this link for an example: \href{https://play.inaf.it/en/how-many-looks-for-a-galaxy/}{https://play.inaf.it/en/how-many-looks-for-a-galaxy/}}}

\subsubsection{Feedback Wheel}
During Phase 1, after the preliminary design of the activities, PhD candidates are asked to deliver their activities to primary school students and/or individuals with BVI. The delivery of the activity is followed by a debriefing aimed at collecting the impressions and feedback of the participants that will be used to modify and improve the workshops. The tool adopted to guide the debrief is the Feedback Wheel (Figure \ref{fig:temp}) developed by the authors. It is a collection of questions about attendees' feelings (how did you feel at the beginning and during the activity?), perceptions (what surprised you? What was difficult for you? Do you think that other children would like this activity?), and criticisms (do you have any other suggestions?) about the activity. Each participant answers by writing on the form and the collected feedback is used to modify the activity.

\subsubsection{ICEBAGS}
\begin{figure}
    \centering
    \includegraphics[width=\textwidth]{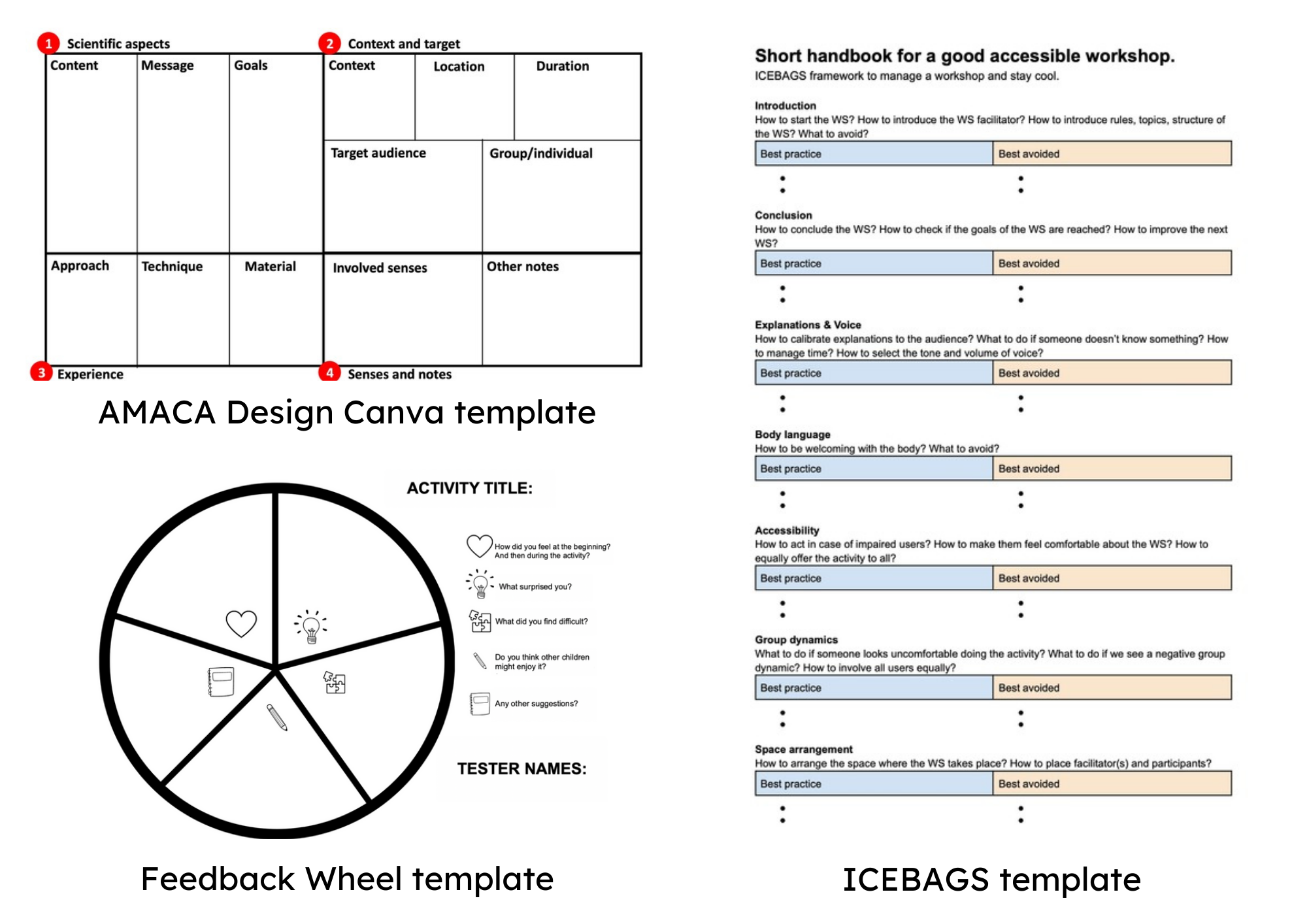}
    \caption{Template of AMACA Design Canvas (upper left), Feedback Wheel (bottom left) and ICEBAGS (right).}
    \label{fig:temp}
\end{figure}

This is a template, developed by the authors, used to create a short handbook for running a good, accessible workshop (Figure \ref{fig:temp}). It is made of six sections, focusing on areas that are key for a successful delivery of hands-on activities, whose initials make the acronym of the template itself: Introduction, Conclusion, Explanations and voice, Body language, Accessibility, Group dynamics, and Space arrangement. For each area, there is a list of best practices and best avoided.
The template is created by PhD candidates as part of the course, during a collaborative lecture (see Table \ref{tab:course}). The filled-in document is then shared with HS students during the training (Phase 2).

\subsection{Materials used for the evaluation of the activities}
\label{subsec:evaluation}

Different evaluation tools and analyses were developed to evaluate the impact of the project, both quantitative (multiple-choice questions) and qualitative (open-ended questions, word clouds, debriefing). Combining the two approaches allowed for a more comprehensive understanding of the outcomes, capturing both measurable trends and deeper insights into participants' experiences that resist quantification \cite{fraenkel1993design}.\\ The project has been approved by the Ethical Committee for the Psychological Research of the University of Padova (code 972-a).

\subsubsection{Questionnaires}
Three questionnaires are proposed to different group targets during the various phases of the project. The goal of all of them is to assess the value and effectiveness of AMACA, as well as possible changes in the awareness of accessibility issues in astronomy, and the potential to discover it through the different sensory channels:
\begin{itemize}
    \item \textbf{PhD candidates:} the questionnaire (presented at the end of Phase 1) aims to assess the impact of the project on their daily lives as researchers;
    
    \item \textbf{HS students and public}: at the end of the Astronomy Festival (Phase 3), the public and HS students are asked to fill out a questionnaire to assess the usefulness and engagement of discovering astronomy in a multi-sensory fashion;
    
    \item \textbf{teachers:} at the end of Phase 4, teachers complete a questionnaire about the training course, its impact on their classes and on their teaching method.
\end{itemize}

The full questionnaires are presented in the Supporting Information section.

\subsubsection{Word clouds}
At the beginning of Phase 2, HS students are asked to answer with one word the question, ``What do you expect from this project?''. At the end of Phase 3 (the Astronomy Festival), they were asked to answer, again with a single word, a different but related question: ``What did you gain from this project?''. 
We collected the answers in two word clouds, which offered a rapid, exploratory way to identify dominant terms and emerging themes in the qualitative data \cite{depaolo2014get}.
The aim was to evaluate whether the expectations and perceptions of the students changed during the project.

\subsubsection{Debriefing}
At the end of the teachers' training program (Phase 4), a debriefing phase is conducted to gather insights about the teachers' experiences. Feedback is specifically sought on the flipped roles between students and teachers, the challenges and benefits of adopting a hands-on approach, and the reactions to teaching science through a multi-sensory methodology. The focus group serves to assess teachers' perceptions and their students' responses to this educational approach.

\section{Results}
\label{sec:results}

In this section we report and discuss the data collected using the tools presented in the previous Section. We divide the results based on the four targets of the AMACA project: Astrophysics PhD candidates, HS students attending the PCTO project, the general public of the Astronomy Festival, and teachers attending the teacher training program.

\subsection{PhD candidates - Phase 1.}
\begin{figure}
    \centering
    \includegraphics[width=\textwidth]{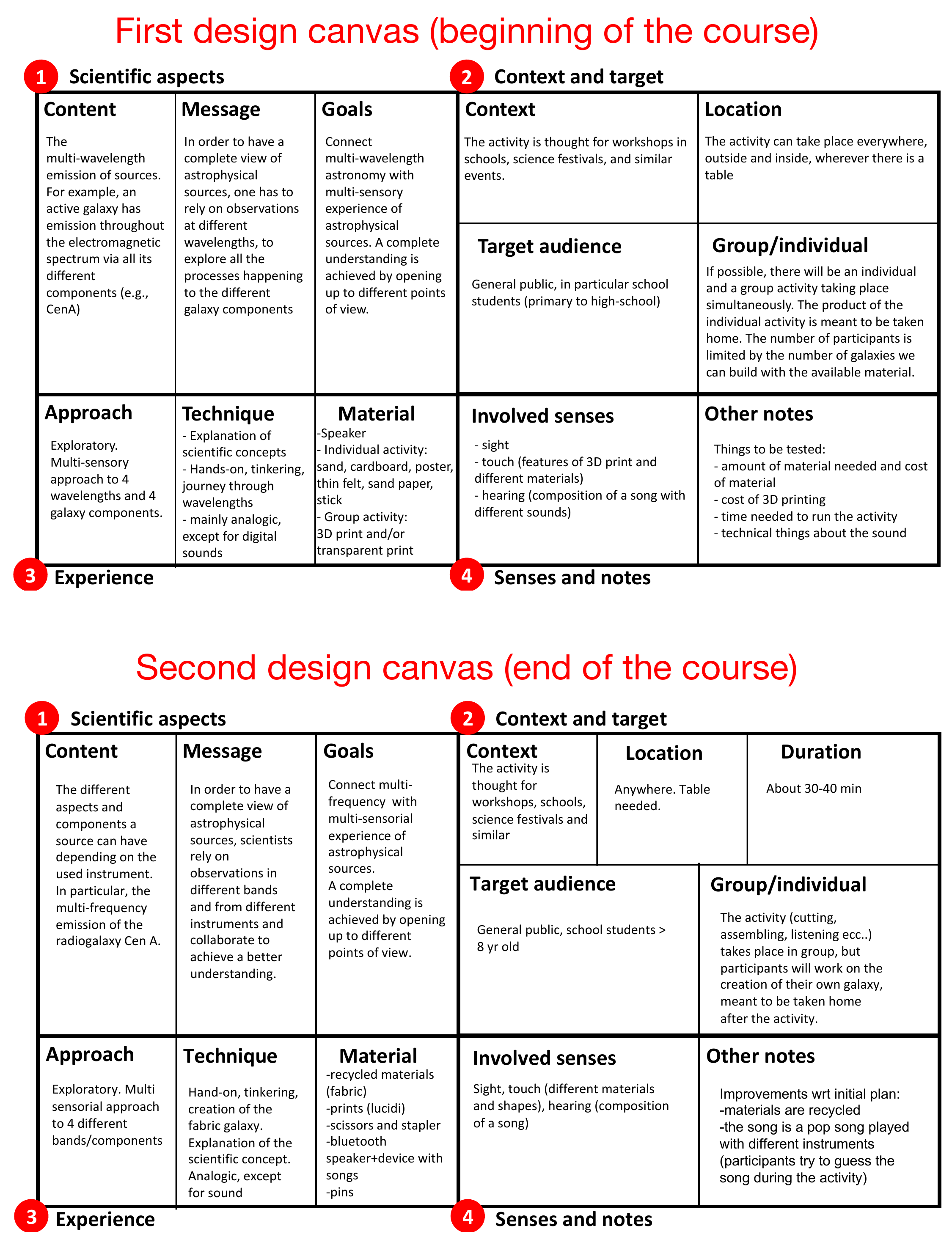}  
    \caption{AMACA Design Canvas filled in by PhD students at the beginning (top) and at the end (bottom) of the course. We show, as an example, the canvas prepared for the activity ``\textbf{The many looks of a galaxy}''.}
    \label{fig:canvas}
\end{figure}

\begin{figure}
    \centering
    \includegraphics[width=\textwidth]{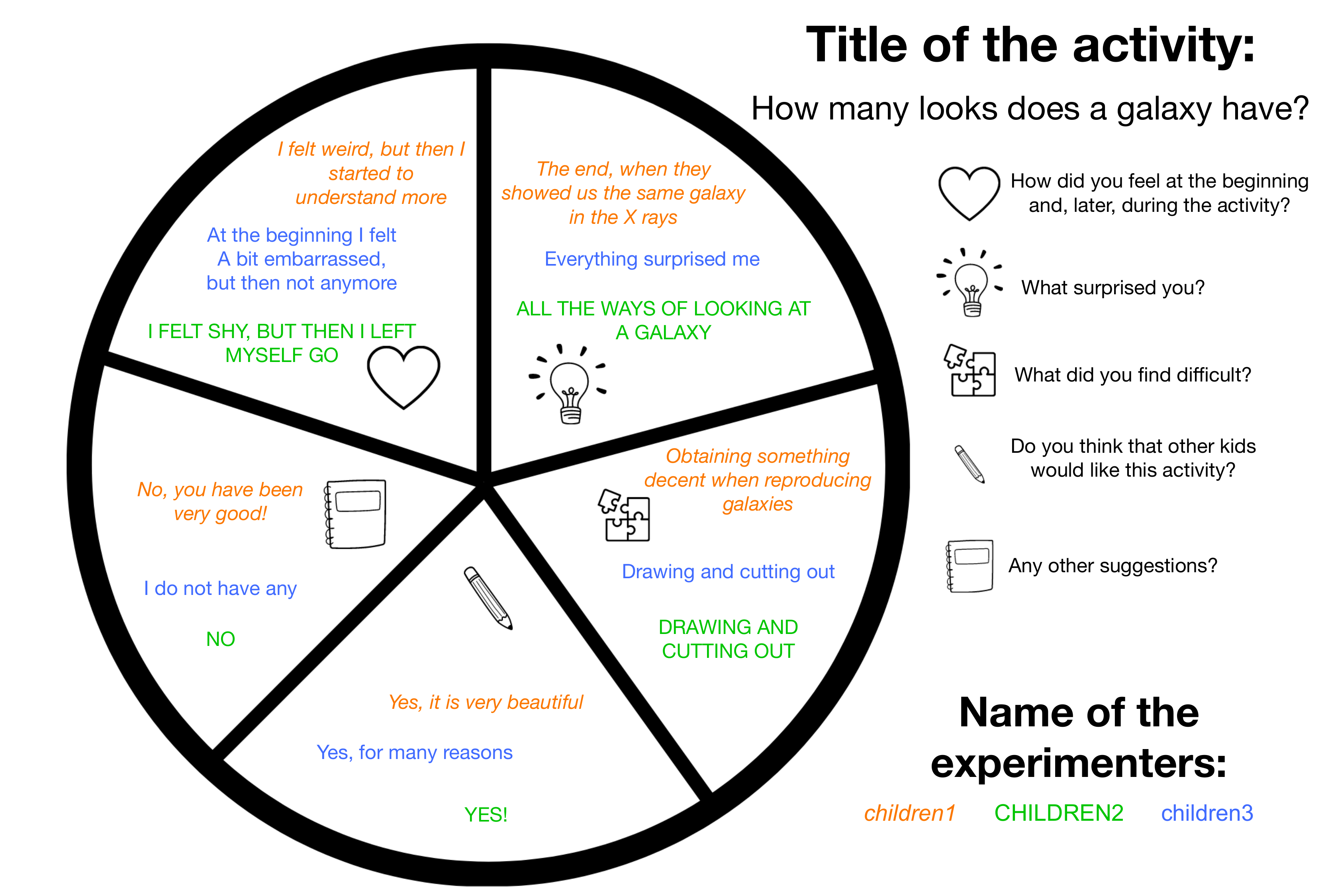}
    \caption{Feedback Wheel obtained by the debrief of the user testing for the activity ``The many looks of a galaxy''}
    \label{fig:debrief}
\end{figure}

The first tool used by PhD candidates is the AMACA Design Canvas, to support the development of the activity. Under the guidance of the lecturers, the candidates complete the Canvas at the beginning of the workshop of Phase 1 and then adjusted using the Feedback Wheel obtained during the pilot workshop with primary-school students. In Figure \ref{fig:canvas} we show, as an example, the two AMACA Design Canvas filled in for the activity ``The many looks of a galaxy''\footnote{\href{https://play.inaf.it/en/how-many-looks-for-a-galaxy/}{https://play.inaf.it/en/how-many-looks-for-a-galaxy/}}, and in Figure \ref{fig:debrief} the Feedback Wheel filled out by three children whose answers are reported with different colors/fonts.

By comparing the AMACA Design Canvas of the activities developed so far, we observed some common changes between the two versions (i.e. those filled in before and after the course), that concern the content, the message and the target audience.

\textbf{Content changes.} In the first design, the statements were vague and encompassed vast astronomical topics, while more specific topics were reported after the trial with primary-school students. This transition is related to the difficulties experienced by PhD candidates, who generally do not have a background in hands-on outreach activities, when presenting general topics to children. Indeed, the approach can be included in the common practice of micro-learning methodology, namely focusing and presenting one single topic at a time to foster meaningful learning \cite{ghafar2023microlearning}.
For example, in the AMACA Design Canvas displayed in Figure \ref{fig:canvas}, the content shifted from ``The multi-wavelength emission of sources'', i.e. the different kinds of light emitted by celestial objects, to ``The different aspects and components a source can have depending on the used instrument''. This feedback-driven change reflects the reaction of participants. Their answers to the Feedback Wheel (Figure \ref{fig:debrief}) showed how they were surprised when they saw the same galaxy in different emission wavelengths, inducing PhD candidates to adjust the content.

\textbf{Message changes.} In the second Canvas, the words ``scientists'' and ``collaboration'' appeared more often than in the first one (as shown in Figure \ref{fig:canvas}), showing an attempt to highlight people and interactions involved in the research world instead of presenting only technologies and results.
Again, this is a common practice in science education and outreach. Astronomy and sciences, in general, are perceived as infallible disciplines conducted by brilliant people who work alone. One of the aims of AMACA is to unhinge this perception and shorten the distance between scientists and the general public by implementing an educational path that bridges professional astronomers, students, and citizens. 

\textbf{Target audience changes.} The ``target audience'' goes through an age-narrowing. In the example of Figure \ref{fig:canvas}, it moved from ``General public, in particular school students (primary to high school)'' to ``General public, school students older than 8 years old''. Indeed, in the Feedback Wheel, two children reported some issues about ``drawing and cutting out'' the materials, leading to the introduction of a lower limit on the age of participants in the second AMACA Design Canvas.

All of these modifications reflect two crucial approaches of AMACA. The first one involves the design of the activities, which includes a testing phase before implementing them in the Festival (Phase 3), using a trial-and-error approach that reflects the scientific method.
The second one, concerns the communication and outreach skills gained by PhD candidates. Nowadays, one crucial part of researchers' duties is to communicate effectively their research, both to the public and experts. This required specific skills that can be achieved through a hands-on approach, as the AMACA project does.

\subsection{High-school students - Phase 2 and 3}
Two word clouds and a questionnaire were used to assess the impact of the AMACA project on HS students. 

\begin{figure}[t!]
    \centering
    \includegraphics[width=0.75\textwidth]{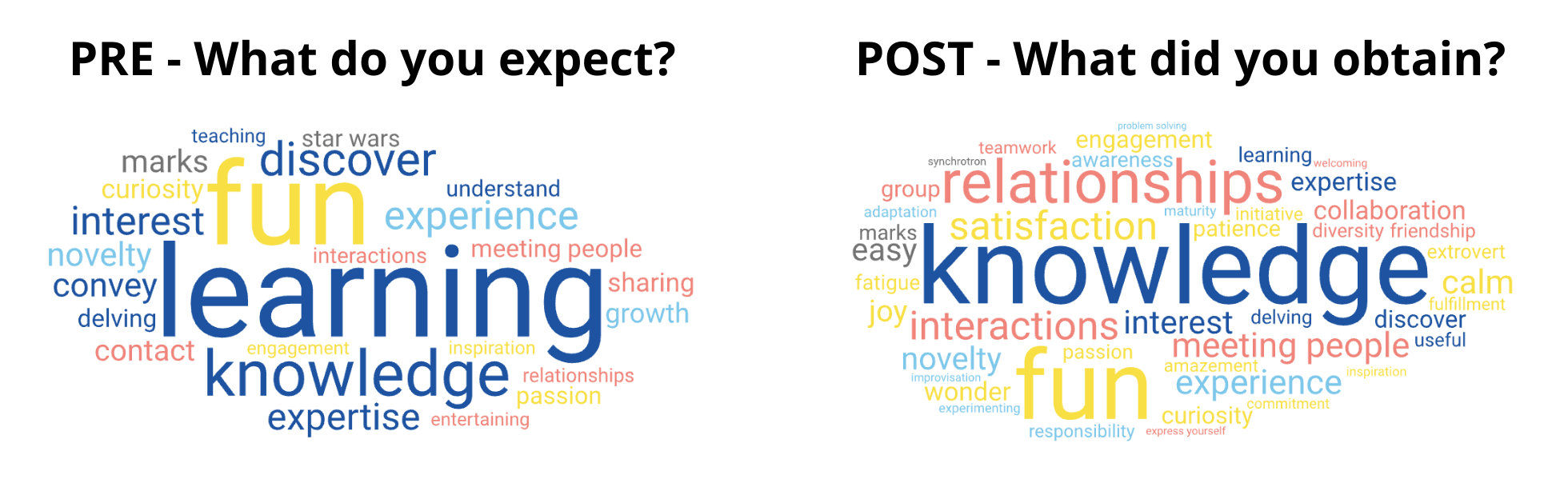}
    \includegraphics[width=0.75\textwidth]{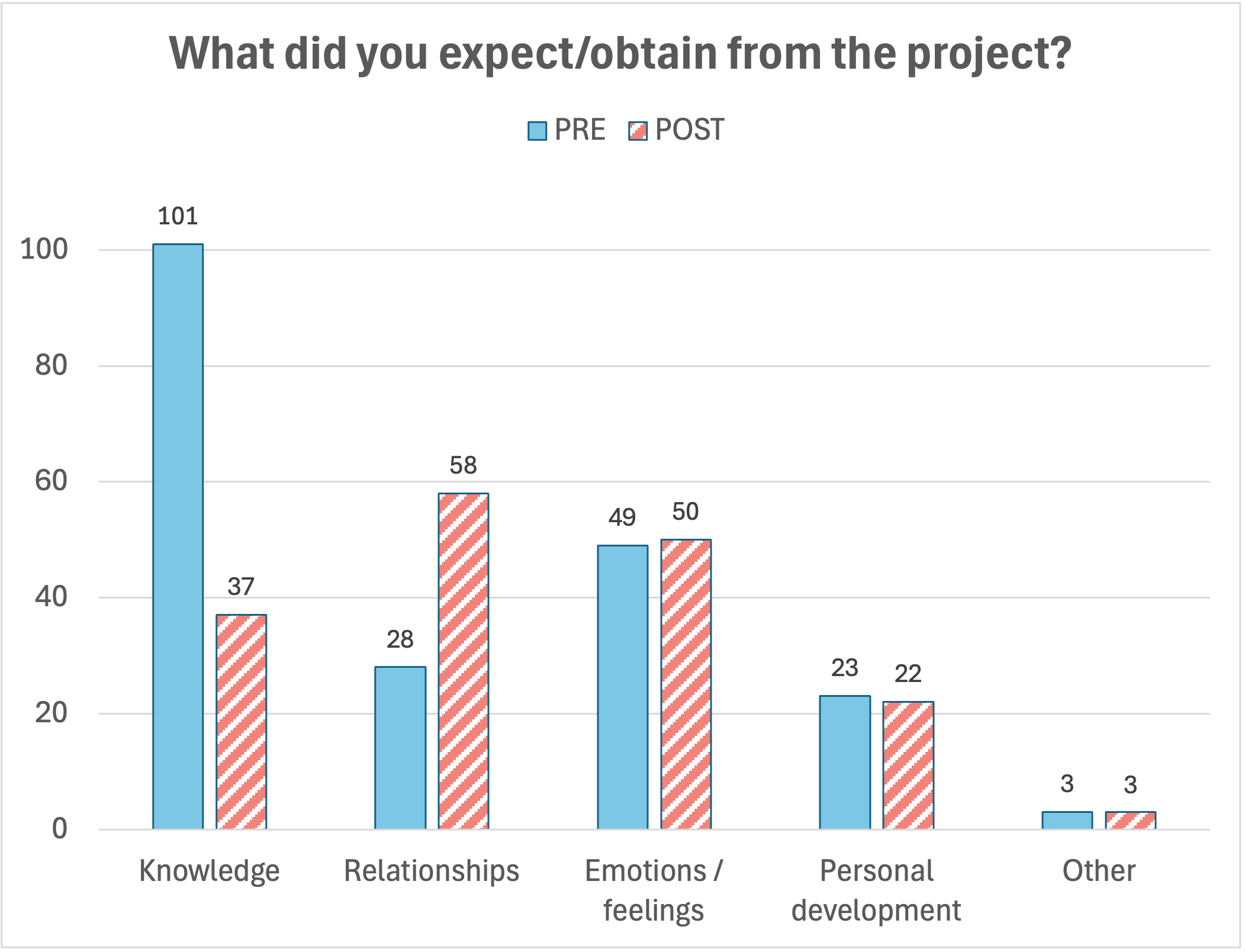}
    \caption{Word clouds generated from HS students' answers to the questions ``What do you expect from this project?'' (left) and ``What did you obtain from this project?'' (right) at the beginning and at the end of the project. Above them are the semantic groups in which the words are represented}
    \label{fig:word_clouds}
\end{figure}

The word clouds reported in Figure \ref{fig:word_clouds} were obtained from 2021 to 2024 and include the answers of 200 HS students aged between 16 and 18 years. They were made with the answers of the students to the questions ``What do you expect from this project?'' and ``What did you obtain from this project?'' respectively at the beginning and at the end of the project. 
To further investigate the comparison of the two word clouds and assess possible changes between students' expectations and achievements, the words were grouped into five semantic groups: knowledge, personal development, emotions/feelings, relationships, and others. The classification was made by running a thematic analysis, which aimed to highlight the main topics addressed by the students. In Table \ref{tab:word_clouds} we report the individual words that were mentioned for each semantic group. The results are reported in Figure \ref{fig:word_clouds}. 
At the beginning of the project, the main expectation is about ``knowledge'' (49.5\%) and the most commonly repeated words are ``learning'', ``knowledge'', and ``discover''. The words related to ``emotions/feelings'' (24.0\%) were put forward (in particular ``fun''), followed by words related to ``relationships'' (13.7\%). That ranking is completely flipped at the end of the project, with ``relationships'' (34.1\%) as the main output, together with ``emotions/feelings'' (29.4\%) and ``knowledge'' (21.8\%). ``Personal development'' has experienced a minor variation (from 11.3\% to 12.9\%).
Looking at the individual words, the number of entries is increased in the second word cloud (from 25 to 45), reflecting the increased complexity and variety of perceptions and feelings experienced during the activities. Most students enrolled in the project because of the astronomical topic, which, in their perception, is related to the rational and educational domain. At the end of the project, they underwent emotions, feelings, and human relationships.
The shift between expectations and achievements is likely due to the high effort required from HS students in terms of relationships and interaction with the public. During the Festival (Phase 3), they are the ones who welcome the public, give them information about the events and manage the workshops, putting into play their problem-solving, communication, and collaboration skills. This actually reflects the daily experience of professional astronomers, who do not carry out their research in isolation only driven by the sake of knowledge, but they rather work in diverse teams where professional relationships, networking, and collaborations are key, they communicate among themselves, and are driven by emotions and feelings such as curiosity, inspiration, fun, passion, while also experience fatigue and frustration, all words that came out in the final word cloud of the HS students.

The second tool that we used to evaluate the impact of the project on the students was a questionnaire about the multi-sensory nature of the activities and the Festival in general. In particular, we asked three questions:
\begin{enumerate}
    \item[\textbf{1.}] On a scale of 1 (= not at all) to 5 (= very much), how engaging did you find it to explore the Universe with all senses? 
    \item[\textbf{2.}] On a scale of 1 (= not at all) to 5 (= very much), how useful did you find it to explore the Universe with all senses for understanding it?
    \item[\textbf{3.}] After attending the Festival, how feasible do you think it is for blind or visually impaired people to study astronomy or work in the field of astronomy? Possible answers to this question were: (1) I am now more confident that astronomy is accessible for blind and visually impaired individuals; (2) I am now less confident that astronomy is accessible for blind and visually impaired individuals; (3) My opinion has not changed.
\end{enumerate}

\begin{figure}
    \centering
    \includegraphics[width=0.75\textwidth]{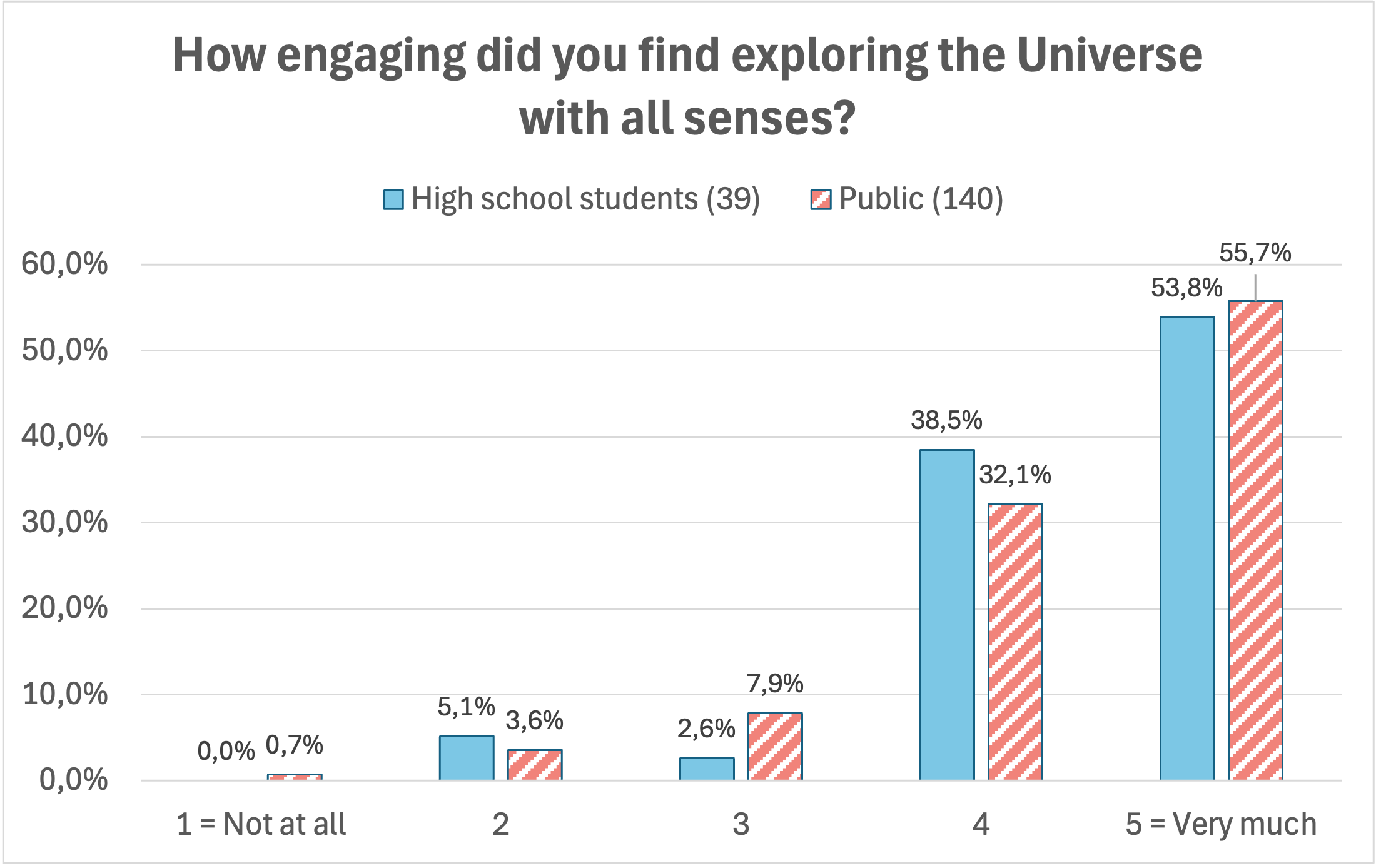}
    \includegraphics[width=0.75\textwidth]{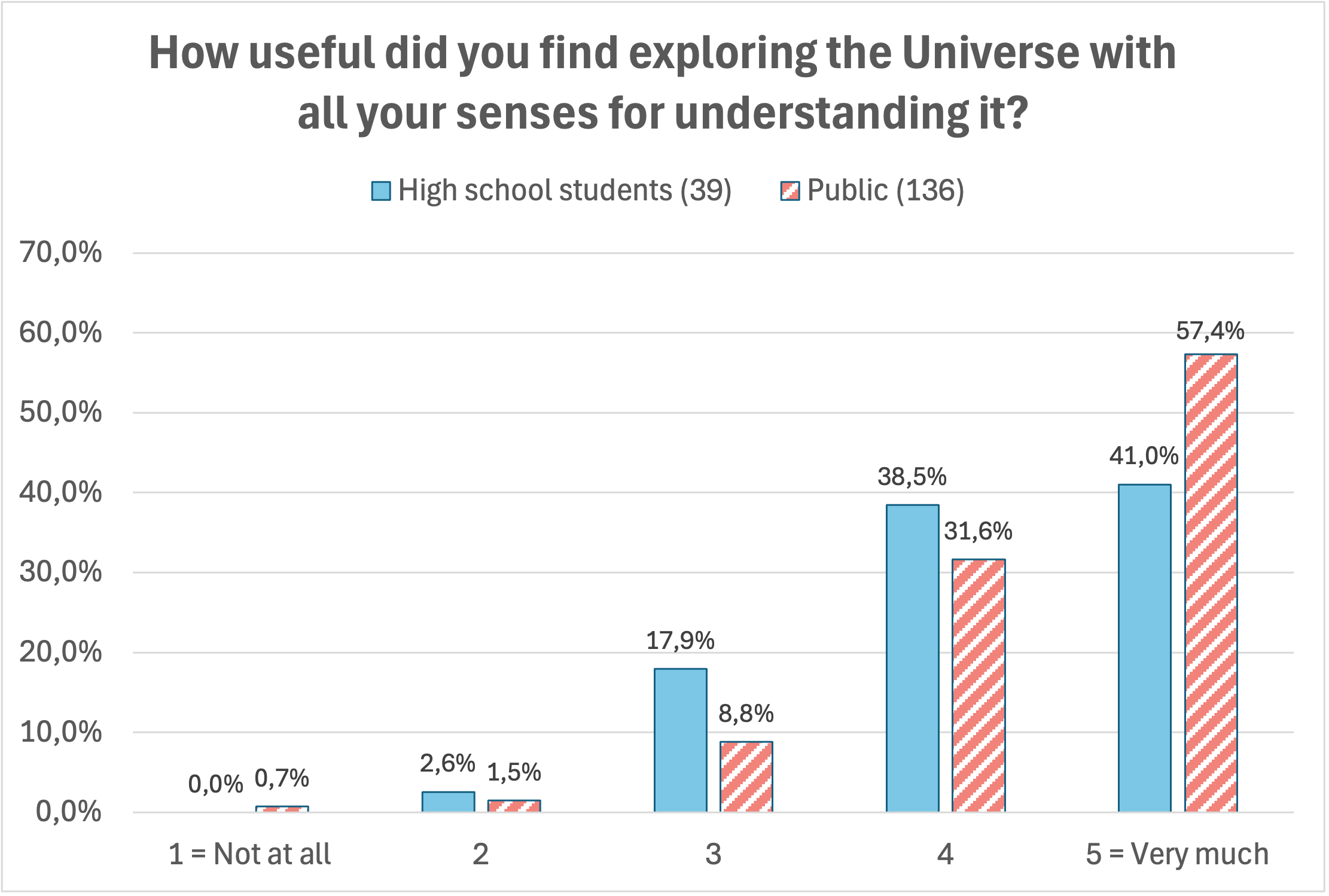}
    \includegraphics[width=0.75\textwidth]{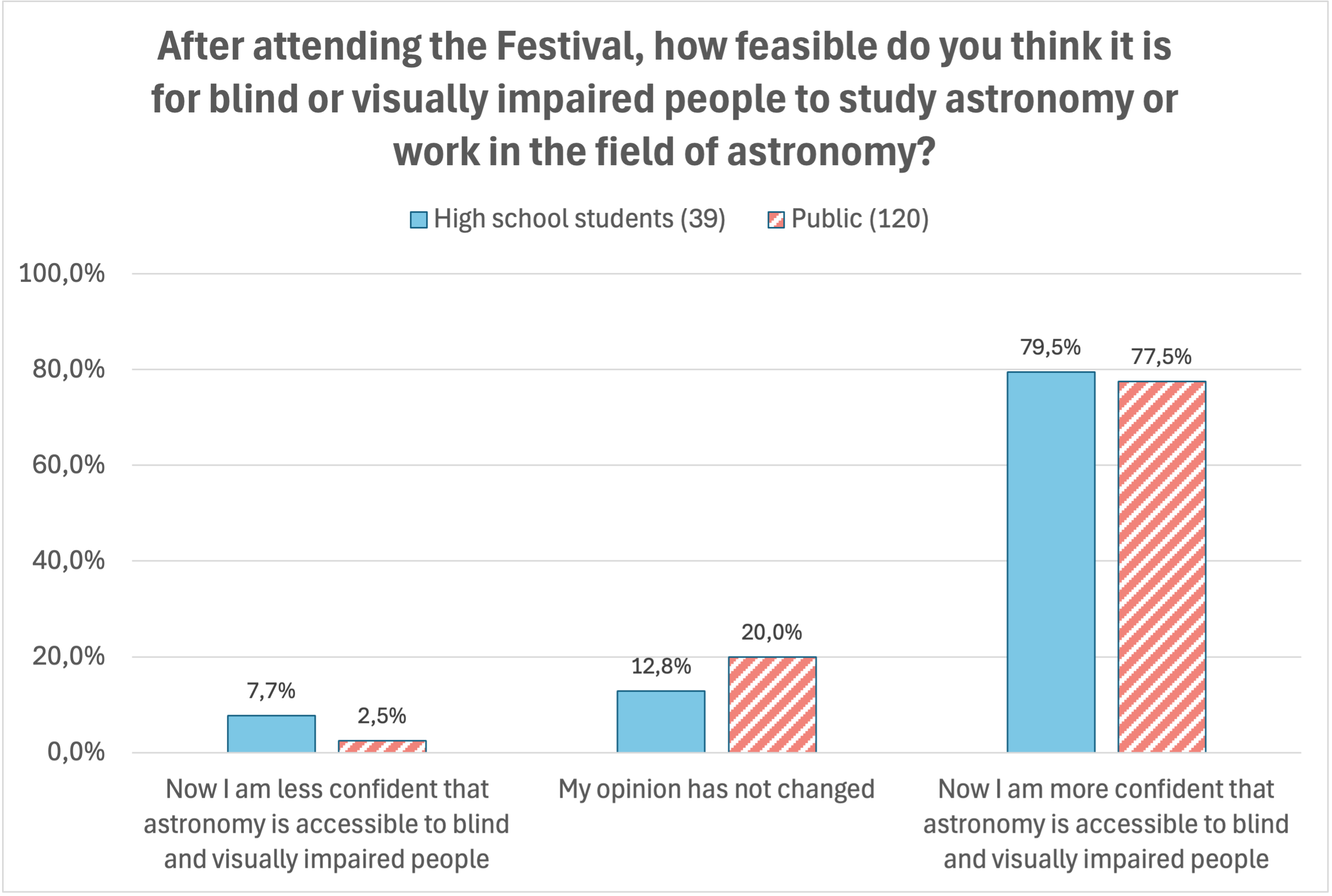}
    \caption{Results of the questions of the questionnaires for each target (HS students and visitors of the Astronomy Festival). Sample sizes are provided in the legend.}
    \label{fig:questionnaires}
\end{figure}

The results of the self-assessment are reported Figure \ref{fig:questionnaires} (blue columns), which includes data of 39 questionnaires collected during 2023. 
Focusing on the first question, most of the students chose 5 (53.8\%) and 4 (38.5\%), indicating that they found the possibility of using multiple senses to learn about astronomy highly engaging. A similar result was obtained for the second question, although with a slightly larger spread (41.0\% for 5, 38.5\% for 4, and 17.9\% for 3). 
The large majority of the students (79.5\%) answered the last question with ``I am now more confident that astronomy is accessible for blind and visually impaired individuals'', followed by ``My opinion has not changed'' (12.8\%).

\subsection{General public - Phase 3}
\begin{quote}
    I want to become an astrophysicist, but I never wondered what blind or deaf people had to do to become astrophysicists and astronomers. Thank you for pointing this out; I had never considered it.\\
   \textit{(10-year-old girl who attended the conference ``The Universe is invisible to the eyes'' held at the Astronomy Festival in 2023\footnote{The video recording is available (in English and Italian) at this link: \href{https://youtu.be/ekjr8sf049U}{https://youtu.be/ekjr8sf049U}})}
\end{quote}
This comment, reported by a 10-year-old girl during the 2023 festival, summarizes better than any other words the potential of the AMACA project. Through the approach described so far, the Festival and the whole educational path can achieve the goal of increasing awareness about multi-sensory astronomy education. That claim is also supported by the data collected through questionnaires reported hereafter.

We asked to the visitors who attended the Festival to fill in a questionnaire including the same three questions asked HS students.
The results are reported in Figure \ref{fig:questionnaires} (orange striped columns), which include data from 141 questionnaires collected in 2023 and 2024. They are comparable to those obtained for the HS students, namely 55.7\% of visitors filling the questionnaires found it ``very engaging'' to learn about astronomy in a multi-sensory fashion, and 57.4\% found multisensoriality useful for a better understanding. Finally, 77.5\% chose the option ``I am now more confident that astronomy is accessible for blind and visually impaired individuals''.

Considering the results of HS students and the general public, the multi-sensory approach appears engaging and useful based on the self-assessment. Additionally, it raises awareness about accessibility and it is effective regardless of the roles (the results from students and visitors are very similar).

An example of the effectiveness of the multi-sensory approach can be found in a comment of a child who attended the activity ``Let's play the stars!''\footnote{The activity is described at this link: \href{https://play.inaf.it/en/8571-2/}{https://play.inaf.it/en/8571-2/}} during the 2023 edition of the Festival. The workshop consisted of the sonification of astronomical images while being blindfolded. Participants moved their hands in front of a computer that displayed an image, that they could not see. Depending on the position of the hand on the digital image, a sound that represents an RGB colour is reproduced - bubbles sound for blue, crackling of the fire for red, and birds chirping for green.
When the facilitator asked the child what colours he perceived from the sounds, he answered, ``I hear blue and red''. This comment suggests that it is easy to change from one sense (hearing) to another (sight) in a matter of seconds.

\subsection{Teachers - Phase 4}
The teacher training ended with a focus group during which participants answered various questions about the course and the workshops in general.
During the discussion, attendees showed high satisfaction with the flipped roles approach, with which HS students become teachers and vice versa. One attendee reported that:
\begin{quote}
    The role of a student is a positive one that challenges you and allows you to learn new ways of doing things and much more.
\end{quote}
For some, this approach led to anxiety, stemming from the fear of not mastering the topic while presenting it to their classes. As one teacher noted, this anxiety ``pushes you out of your comfort zone'', resulting in greater empathy for students who often experience similar emotions in their daily school life.

To teachers, HS students appeared motivated and engaged, even if they would not have been evaluated with scholastic marks. This was a positive surprise for the teachers, who were prone to think that, without the pressure of being graded, students would not be prepared, as reported in the following teacher comment:
\begin{quote}
    I do not know if students get some kind of recognition or not for participating, but when there are no marks, it is hard for someone to really give their best. I was very positively impressed. (...) I was very surprised to see them perform so well, even without an immediate benefit.
\end{quote}
Moreover, they valued the interactive and hands-on approach of the course that showed them the workshops in action rather than being simply a theoretical training:

\begin{quote}
    Having seen the workshops at the festival, then taking the course, and later trying to manage them myself, I realized their potential and strength, especially by doing it.
\end{quote}
Thanks to the experience acquired during the Festival, HS students highlighted the strengths and weaknesses of each activity and gave suggestions for implementing the activities effectively in the classrooms. As a teacher reported:
\begin{quote}
    When I managed the workshop on my own, I thought about them [HS students], what they had asked me, and what they had told me.
\end{quote}
The hands-on and multi-sensory approach of the activities has also been appreciated, confirming its ability to engage and motivate even the most disillusioned students:
\begin{quote}
    I also found the workshop very positive. It was in a very challenging class with many people who are disillusioned with school and quite nihilistic. They all participated in their own way; it was engaging and exciting for them.
\end{quote}
\begin{quote}
    These activities greatly enrich the lesson because they use different methods compared to just a traditional lecture. The students are more engaged, so these labs are very welcome.
\end{quote}

\section{Discussion}

The primary goal of AMACA is to raise awareness about the potential of a multi-sensory approach to astronomy, a topic that has gained interest and been studied in various projects in recent years (see, for example, \cite{Perez-Montero2019, Noel-Storr2022, Foran2022, Harrison2022, GuiottoNaiFovino2023, Varano2023, Varano2024}). As stated by \cite{Harrison2023}, there are several limitations preventing the widespread adoption of these resources and methods, such as: (1) many have been developed for one-time events; (2) the materials used are often expensive; and (3) they require a facilitator with expertise in the subject. What AMACA contributes to this field is an approach that leverages multi-disciplinary collaboration between experts in astronomy education, psychologists specializing in educational psychology and disabilities, and representatives from the Italian Union of the Blind and Visually Impaired as well as the Italian National Deaf Organization. This collaboration uses a Universal Design for Learning (UDL) approach to co-design hands-on, accessible, and affordable educational activities in a recurring annual effort that reaches various audiences in both formal and informal settings. Additionally, the development of an educational pathway removes the need for astronomy experts to conduct the activities. Finally, all resources have been made freely available in the INAF repository PlayINAF (\href{https://play.inaf.it/}{play.inaf.it}).

In the following, we discuss the lessons learned during the first four years of the AMACA project, we report some guidelines and recommendations, and highlight criticalities to be addressed in the future. 

\subsection{Lesson learned}

\begin{enumerate}
    \item \textbf{Multi-sensory approaches change the perception of astronomy as a mono-sensory science}. The evaluations that we conducted with HS students and visitors participating in the Astronomy Festival ``The Universe in All Senses'' have shown AMACA's ability to challenge the perception of astronomy as a mono-sensory science. The results shown in Figure \ref{fig:questionnaires} demonstrate that both target groups found the multi-sensory exploration to be engaging and useful, and also changed their perceptions regarding the feasibility of an individual with BVI pursuing a career in astronomy. These outcomes can be attributed to the multi-sensory nature of the activities experienced by the public and HS students during the Festival. As stated by \cite{perez2024astroaccesible}, the UDL approach not only benefits everyone, but it also potentially leads to the professional involvement of individuals with BVI. Indeed, during the past editions, some of the HS students who got involved in the project were visually impaired. This has indeed been observed in astronomy projects such as Audio Universe, a multi-sensory planetarium show, which has proven to benefit not only individuals with BVI, but also sight ones \cite{Harrison2023}.

    \item \textbf{Community involvement enhances inclusion and reach} 
The limited duration of festivals (typically 3-4 days) intensifies participants' sense of investment and excitement and the interaction with experts leading the activities has been shown to play a crucial role in generating positive impacts on visitors \cite{jensen2014people}. Additionally, one of the key benefits of festivals is the possibility for the visitors to ``identify their own individual `path' to engaging with the scientific content'' \cite{bultitude2014science}. On the other hand, as \cite{kennedy2018preaching} pointed out, science festivals predominantly attract ``economically privileged and educated audiences already invested in science''. To address this issue, \cite{bultitude2014science} suggested holding events in unconventional locations and incorporating cross-disciplinary activities. These approaches can help broaden the audience by engaging people who may not initially be interested in science. AMACA adopts both strategies. We chose to hold the Festival in Castellaro Lagusello, a small village in northern Italy known primarily for its medieval history and beauty. Local residents participated by offering their gardens and courtyards as venues for the activities. This level of community involvement would not have been possible had the Festival taken place in a large city, as is the case with most outreach festivals.
Also the involvement of HS students aimed to engage all, including those who were not strictly interested in astrophysics or scientific subjects in the first place. In fact, several students participated in the project out of necessity, as they were required to earn credits to complete the academic year.
   
    \item \textbf{Multi-disciplinary activities broaden participation}. To engage a broad audience beyond just those interested in science, we employed a crossover of disciplines by developing several activities that combine themes from both astrophysics and art, such as astronomical poetry and wall mapping of stunning astronomical images set to music. Also using a multi-sensory approach increased public interest due to the novelty and originality of the experience. Additionally, the multidisciplinary nature of the activities also had an impact on teacher training, attracting not only science teachers but also educators from various disciplines (e.g., an Italian language professor, a special education teacher, a history teacher).

    Finally, both the hands-on and the multi-disciplinary approach helped us engage a larger number of students in educational contexts than more traditional science classes \cite{freeman2014active}. This was highlighted by the teachers participating in the training, who also appreciated the flipped roles of students and teachers during the course. We will explore this latter aspect in greater depth in a future study.

    \item \textbf{Flipping the roles of ``experts'' and ``audience'' shifts perceptions of astronomy}. The interaction between astrophysics PhD candidates and HS students fostered a better understanding of the target audience and how to effectively communicate science for the former, and provided the latter with a broader and more nuanced understanding of astronomy. Additionally, the role reversal — where the ``audience'' becomes ``experts'' (e.g., PhD candidates who initially took the course later train HS students, and those students, after the training, become the ``experts'' both during the Festival and in teacher training) — enhanced the responsibility and engagement of all participants. This emotional involvement and sense of responsibility played a key role in expanding their perception of the subject, fostering both openness and curiosity. In fact, as shown in Figure \ref{fig:word_clouds}, HS students involved in the project experienced a shift in their perception of what astronomy actually is, moving from educational-domain expectations at the beginning of the project to a more complex understanding that includes emotions and interactions in their learning experiences. The emergence of positive emotions temporarily broadens cognitive and physical resources, fostering openness in thinking and planning, which can lead to a state of curiosity that promotes autonomous learning \cite{fredrickson2004broaden_emozionipositive}.
This aspect will be further investigated in future \textit{ad hoc} studies.

Similarly, teachers appreciated the flipped roles during training sessions, which empowered them to take an active part in their learning and to develop empathy with their students. Also this aspect will be further investigated in the future. 
\end{enumerate}

\subsection{Guidelines and recommendations}

\begin{enumerate}
    \item \textbf{Adopt UDL and co-design framework}. The multi-sensory nature of the educational activities and the UDL approach have to be part of the design since the initial, conceptual phases. Involving experts in education, psychology, sound design, and representatives from disability organizations during the co-design of the activities ensures their effectiveness with different targets.
    \item \textbf{Leverage festivals for concentrated impact}. Organize events with a defined duration (e.g., 2-3 days) to generate excitement and encourage focused participation.
    \item \textbf{Choose unconventional locations}. Select small towns or unique venues to create a welcoming and inclusive atmosphere. Engage the local community in the organization and execution of the event to expand the audience.
    \item \textbf{Incorporate multi-disciplinary activities}. Blend science with other disciplines like art, music, or literature to appeal to diverse audiences and increase engagement.
    \item \textbf{Make cheap resources freely available}. Provide all materials online in an open-access repository to ensure widespread use and adoption by educators and teachers. Use cheap materials to increase dissemination.
    \item \textbf{Encourage active participation}. Design activities that allow students to take an active role in teaching or facilitating. This empowers them, improves their understanding, and fosters their enthusiasm for science.
    \item \textbf{Engage broader audiences}. Develop strategies to attract people outside the typical “science-engaged” demographic. Combine traditional science communication with activities that appeal to diverse interests, and hold events in settings that reach underrepresented groups.
    \item \textbf{Exploit educational paths that involve different groups.} Developing activities that bring together different target groups - such as students, teachers, and astronomy experts - promotes the spread of awareness and provides opportunities of interaction. Involving participants with diverse abilities also enhances empathy, understanding, and cooperation.
\end{enumerate}

\subsection{Challenges}

\begin{enumerate}
\item \textbf{Time investment and professionality.} The organization of the AMACA educational path requires a significant amount of time and constant efforts throughout the year, with tight deadlines and a high level of commitment. Additionally, due to the diverse range of tasks and aspects the project involves, organizers with varied professional backgrounds are essential. The AMACA team currently includes astronomers, teachers, disability experts, psychologists, an architect, and a website designer. However, to ensure the project's long-term sustainability and enhance its professionalism, additional expertise is needed, particularly in fundraising and communication.

\item \textbf{Accessibility of venues and events}. Although unconventional locations can broaden participation, ensuring that venues are accessible can sometimes be a challenge. For example, some venues may not be served by public transportation, limiting access to those with cars. Furthermore, ensuring the accessibility of historical sites, such as the medieval village of Castellaro Lagusello, or private locations like the gardens opened by local residents for the Festival, remains an ongoing issue. In collaboration with the local administration, we are exploring options to organize shuttle services from the nearest train stations to Castellaro Lagusello for the entire duration of the Astronomy Festival. Additionally, we are providing recommendations to the administration on how to improve the accessibility of public spaces for people with disabilities. These suggestions, which will enhance the accessibility of the Festival, will also have a lasting impact, benefiting the community throughout the year and setting an example for other small towns.

\item \textbf{Reaching underrepresented groups}. Astronomy is often perceived as a visual science, which can make it challenging to engage audiences, such as the visually impaired, who are typically excluded from astronomy education and outreach. To address this, we are collaborating with local organizations for blind and deaf individuals to raise awareness about the Astronomy Festival and the educational resources developed within the AMACA framework. We are also working with schools to actively encourage the participation of students with disabilities in both the high-school training program and the Festival itself.

\item \textbf{Test the efficacy}. The multi-sensory approach developed in AMACA has proven to be engaging and enjoyable. Further research should assess its effectiveness in promoting a deeper understanding of astronomy and supporting effective learning.
\end{enumerate}

\section{Summary and conclusions}
\label{sec:conclusions}
In recent years, innovative approaches have been developed to make astronomy communication multi-sensory. While promising, these efforts face challenges that prevent widespread adoption. Many resources are created for one-time events, poorly documented, or difficult to share. Tactile materials are often expensive and hard to produce on a large scale. Most initiatives are developed by astronomers alone, without interdisciplinary collaboration, and lack user testing. Additionally, dissemination is usually limited to the original creators, making wider use challenging.

To address these issues, we developed AMACA (``Astronomy education with a Multi-sensory, Accessible, and Circular Approach''), aiming to promote multi-sensory astronomy through education and public engagement. AMACA’s educational path consists of four phases:

\begin{itemize}
    \item \textbf{Phase 1}: University course for Astrophysics PhD candidates, focusing on creating multi-sensory outreach activities. Lectures are given by astronomers and psychologists, with input from the Italian Union of the Blind and Visually Impaired (UICI).
    
    \item \textbf{Phase 2}: High-school training. PhD candidates teach HS students the activities they designed, with support from UICI and ENS.
    
    \item \textbf{Phase 3}: Astronomy Festival. HS students lead hands-on workshops and engage the public at ``The Universe in All Senses'' festival.
    
    \item \textbf{Phase 4}: course for teachers of all school cycles (kindergarten, primary, middle, and the first years of high school). HS students and astronomers  train teachers about the hands-on activities.
\end{itemize}

We used tools to guide activity development (e.g., AMACA Design canvas, ICEBAGS), feedback mechanisms (e.g., Feedback Wheel, Word clouds), and questionnaires to assess impact (Fig. \ref{fig:temp}).

The key results of our study are:
\begin{itemize}
    \item \textbf{PhD candidates}: By analyzing the AMACA Design Canvas filled out by PhD candidates at the start and end of the course (Phase 1), we found that by the end, students had sharpened the focus of their activities, clarified the message they wanted to convey, emphasized people and interactions in the research process (rather than just technology and results), and narrowed the target audience by age. This demonstrates that the students improved their skills in understanding their audience and communicating research. 

    \item \textbf{HS students}: questionnaires and word clouds revealed that while HS students initially associated astronomy mainly with ``knowledge'' and ``learning'', by the end of the project, they began to connect it with emotions, feelings (e.g., ``fun''), and relationships. Most students also found the use of multiple senses in learning astronomy to be highly engaging and beneficial for their understanding.

    \item \textbf{General public}: We asked the public how engaging and helpful the use of all senses was in understanding astronomical concepts. The results were similar to those of the HS students. Additionally, 77.5\% of visitors expressed greater confidence that astronomy is accessible for blind and visually impaired individuals.

    \item \textbf{Teachers}: The teacher training concluded with a focus group, where teachers expressed high satisfaction with the flipped roles approach, in which HS students became teachers and vice versa. They also appreciated the interactive, hands-on nature of the course, which allowed them to see the workshops in action, rather than just receiving theoretical training.
\end{itemize}

Overall, AMACA’s multi-sensory, hands-on approach benefited PhD candidates, HS students, the general public, and teachers, making astronomy more engaging and accessible for all.

\section*{Acknowledgements}

We acknowledge the contribution of F. Di Giacomo and S. Sandrelli in all phases of the project. We thank B. Arfè, L. Martini, C. Boccato, and S. Varano who contributed to the PhD course. We are grateful to UICI and ENS representatives who contributed to the training of the high-school students. We thank the PhD students, teachers, and high-school students who made the AMACA educational path possible. We thank G. Tirabassi and L. Guiotto Nai Fovino for the discussions about the evaluation of the AMACA project. We thank IAU Office of Astronomy for Education Center Italy for the support.

\section*{Disclosure statement}

The authors report there are no competing interests to declare.

\section*{Funding}

This work was supported by the INAF grant ``An inclusive educational chain culminating in the Astronomy Festival'' (Ob. Fu. 1.05.17).

\section*{Ethics Statement}
The project has been approved by the Ethical Committee for the Psychological Research of the University of Padova (code 972-a).

\bibliography{bibliography}

\bigskip


\appendix

\begin{landscape}

\section{Courses program}
\label{supporting_info}

\subsection{Course for PhD candidates}
The course for Astrophysics PhD candidates held at the University of Padova and at the University of Bologna has a duration of 18 hours. Lectures are generally performed in blocks of two hours each, back-to-back. In Table \ref{tab:course} we report an outline of the course. Participants are strongly encouraged to explore these subjects further on their own. To support this, we provide references to academic papers, relevant books, and online resources.

\begin{longtable}[t!]{p{1.7cm}| p{8.5cm}| p{1.5cm}| p{7cm}}
  \caption{Outline of the course for Astrophysics PhD candidates.}\\
    \hline
    \textbf{Lecture}   & \textbf{Topic} & \textbf{Duration} & \textbf{Learning objectives} \\
    \hline
    Lecture 1 (February)    &  Introduction of the course (structure, goals, schedule). & 30 min & Familiarize with the concepts of communication, outreach, public engagement, education. \\
                & Introductory activity about communication and public engagement. & 30 min & \\
    \hline
    Lecture 2 (February) & Motivation for a multi-sensory approach to research, public engagement, and education. & 1 h & Learn about tactile and sonification approaches. \\
    \hline
    Lecture 3 (February) & Hands-on activity with the Italian Union for the Blind and Visually Impaired (UICI) about visual disabilities. & 1 h & Familiarize with some of the challenges of visual impairments.\\
    \hline
    Lecture 4 (February) & Hands-on activity with the responsible of the Padova University for Inclusion and Diversity about auditory disabilities. & 1 h & Familiarize with some of the challenges of auditory impairments. \\
    \hline
    Lecture 5 (February) & Participation in two multi-sensory, hands-on activities developed by AMACA in previous years. Participants can decide to participate blindfolded to experience the lack of sight. & 1 h & Familiarize with the hands-on approach to science communication.\\
    \hline
    Lecture 6 (February) & Debrief using ICEBAG to reflect on how the activities were facilitated. & 45 min & Understand how to communicate scientific subjects to the general public verbally and with body language, and how to engage the audience.\\
              & Presentation of topics they can focus on to develop their activities (e.g., Moon phases; sundials; entropy; ...). & 15 min & \\
    \underline{At home} & The participants decide if they want to work alone or in couple (and with whom). They are asked to choose the topic they want to focus on. & & \\
    \hline
    Lecture 7 (March) & Activity about imagining and describing targets (e.g., a 7 year-old child, a blind 50 year-old adult, a 14 year-old teenager, ...) to understand what are the key characteristics of the activities tuned for a specific target. & 1 h & Empathize with the target audience to understand their needs/knowledge. \\
    \hline
    Lecture 8 (March) & Presentation of the AMACA Design Canvas. & 15 min & \\
            & The participants fill in their canvas with their idea of activity. & 30 min & Learn to use a design tool and to create the concept of educational activities.\\
           & The groups present their canvas plenary and ideas are exchanged. & 1.25 h &\\
    \underline{At home} & The participants are asked to work on the canvas and implement the suggestions that were given to them during the lecture. They are also asked to come up with a title and a minimum age for the target of their activity. &  & \\
\hline
    Lecture 9 (March) & They are presented the tinkering activity ``Build your roller coaster!'' (\href{https://play.inaf.it/montagne-russe/}{link}) & 30 min & Learn about the tinkering approach. \\
              & Discussion about tinkering (origin, concept,  bibliography, ...) & 30 min & \\
            & Guiding the design of the prototypes (in groups). & 1 h & \\
\hline
Lecture 10 (March) & Guiding the design of the prototypes (in groups). & 1 h & Learn how to prototype educational activities. \\
\underline{At home} & The participants are asked to gather the material needed to build and start creating the prototypes. & & \\
\hline
Lecture 11 (April) & Activity about storytelling & 30 min & Learn about the storytelling approach. \\
           & Discussion about storytelling (concept, techniques, bibliography, ...) & 30 min &  \\

            & Guiding the design of the prototypes (in groups). & 1 h & \\
\hline
Lecture 12 (April) & Guiding the design of the prototypes (in groups) & 1 h & Learn how to document educational activities. \\
\underline{At home} & The participants are asked to finalize the activities and start writing a report about the activities. & & \\
\hline
Lecture 13 (April) & Finalizing the activities (in groups) & 1 h & Learn how to prototype activities until the first design is complete.\\
\hline
Lecture 14 (April) & Finalizing the activities (in groups) & 1 h & Learn how to prototype activities until the first design is complete.\\
\underline{At home} & The participants are asked to practice their activities to be ready to facilitate them with children and users of the target age. & & \\
\hline
Lecture 15 (May) & Testing the activities with children and users of suitable age. & 1 h & Learn how to run the activities with actual users. \\
\hline
Lecture 16 (May) & Debrief with children and users using the Feedback Wheel & 30 min &  Learn how to collect users' feedback. \\
     & Debrief with PhD candidates & 30 min &  \\
\underline{At home} & The participants are asked to think how to improve the activities based on the feedback received from the children. They are asked to fill in the AMACA Design Canvas again. & & \\
\hline
Lecture 17 (May) & Presentation to everyone of the new canvas and finalization of the activities.  & 1 h & Learn how to modify the design to implement users' feedback.\\
\hline
Lecture 18 (May) & Presentation to everyone the new canvas and finalization of the activities. & 30 min & Learn how to document the activities.\\
          & Wrap up of the course. & 30 min & \\
\underline{At home} & The participants are asked to update the report of the activities and prepare their final version. They are also asked to practice the activities to be ready to transfer the knowledge to HS students. & & \\
\hline
Final exam (May) & Knowledge transfer to HS students. & 4 h & Learn how to train educators.  \label{tab:course}
\\
\hline
\end{longtable}

\subsection{Course for HS students}
The course for HS students involves both a first part of training and a second part to put in practice the acquired knowledge. We report an outline of the course in Table \ref{tab:training_highschool}.

\begin{longtable}{p{1.7cm}| p{8.5cm}| p{2cm}| p{7cm}}
    \caption{Outline of the course for HS students.}\\
    \hline
    \textbf{Lecture} & \textbf{Topic} & \textbf{Duration} & \textbf{Learning objectives} \\
    \hline
    Lecture 1 (May) & Introduction of the course and icebreaker activities. & 30 min & Acquire the astronomical background needed for the activities. \\
        & Introduction of the astronomy topics that are the subject of the activities developed by PhD candidates. & 2 h (plus 15 min break) & \\
        & Question and answers. & 15 min & \\
         & HS students are divided in small groups and are trained about the activities developed by PhD candidates. & 1.5 h & \\
    \underline{At home} & Students are asked to study the material given during lecture 1 about the astronomical topics and the form prepared by PhD candidates about the specific activity that each group has chosen. & & \\
    \hline
    Lecture 2 (May) & Hands-on activity with the UICI about visual disabilities. & 1.5 h & Familiarize with some of the challenges of visual and auditory impairments. Learn how to welcome and engage the public.\\
             & Hands-on activity with the ENS about auditory
disabilities. & 1.5 h & \\
    \underline{At home} & Students are asked to study the ICEBAG and reflect on the inputs obtained during the activities with UICI and ENS. & & \\
    \hline
    Lecture 3 (June) & HS students are asked to rehearse the activities that they will present during the Festival. They perform several rehearsals and receive feedback each time until a high-level presentation is reached. & 3 h & Learn to deliver the activities. \\
    \hline
    Lecture 4 (June) & Facilitation of the activities during the Astronomy Festival. & 11 h (over 2 days) & Deliver the activities to the public. \\
    \hline
    Lecture 5 (October) & Refresh memory: HS students are asked to rehearse the activities that they presented during the Festival. & 3 h & Remember the content and communication methods for the activities.\\
    \hline
    Lecture 6 (October) & HS students train teachers about the activities. & 3 h & Deliver the activities to teachers. \label{tab:training_highschool}
\\
    \hline
\end{longtable}

\newpage
\subsection{Teacher training}
The teacher training lasts for 10 hours and is held between October and December. In Table \ref{tab:course_teachers} we outline the programme of the course. Participants are strongly encouraged to explore these subjects further on their own. To support this, we provide references to academic papers, relevant books, and online resources.

\begin{longtable}{p{1.7cm}| p{8.5cm}| p{2cm}| p{7cm}}
    \caption{Outline of the course for primary- and secondary-school teachers.}\\
    \hline
    \textbf{Lecture} & \textbf{Topic} & \textbf{Duration} & \textbf{Learning objectives} \\
    \hline
    Lecture 1 (October) & HS students train teachers about the activities. Each teacher is trained on three different activities. & 3 h & Learn the astronomy content of the activities and the hands-on approach. \\
        & Debrief and calendar of next lectures. & 1.5 h & \\
    \hline
    Lecture 2 (October) & One of the astronomers supervising the whole AMACA project leads one of the activities chosen by teach in his/her classroom. This is repeated for each teacher. & 1 h (per teacher) & Learn how to deliver hands-on activities to a class. \\
        & Debrief. & 30 min (per teacher) &\\
    \hline
    Lecture 3 (November) & Each teach leads one of the chosen activities in his/her classroom, with the supervision of one astronomer. & 1 h (per teacher) & Learn how to deliver hands-on activities to a class. \\
        & Debrief. & 30 min (per teacher) & \\
    \hline
    Lecture 4 (December) & Questions and answers. Debrief. & 2.5 h & \label{tab:course_teachers}
\\
    \hline
\end{longtable}
\end{landscape}

\appendix
\renewcommand{\thesection}{\Alph{section}}  
\setcounter{section}{1}
\section{Materials}

All the materials presented in this Section are available at the OSF Repository (\href{https://osf.io/7q69z/}{https://osf.io/7q69z/}).

\subsection{AMACA Design Canvas}
\label{subsec:canvas}

The Design Canvas (Figure \ref{fig:temp}) consists of a series of squares to be filled with the main characteristics of the activity, that are:
\\
\noindent \textbf{1. Scientific aspects}:
\begin{itemize}
\item \textbf{Content}: what scientific topic do you want to focus on? What is interesting for you to transmit, and what could be interesting for the audience to learn?
\item \textbf{Message}: what message do you want to convey? What would you like people to learn, understand, and remember? 
\item \textbf{Goals}: what do you want to achieve? Why should the audience come and care?
\end{itemize}

\noindent \textbf{2. Content and target}
\begin{itemize}
\item \textbf{Context}: where will the activity take place? In a classroom, science festival, research conference, planetarium?
\item \textbf{Location}: will the activity be inside or outside? What kind of space is needed?
\item \textbf{Duration}: how long is the activity meant to last?
\item \textbf{Target audience}: who will be the main involved audience? Adults, kids? With disability? Teachers, researchers, general public?
\item \textbf{Group/individual}: is this an activity done in group or individually? What is the optimal group size? Is there a maximum number of participants to the activity?
\end{itemize}

\noindent \textbf{Experience}
\begin{itemize}
\item \textbf{Approach}: what approach will you take? A ‘lecture’ approach or an ‘exploratory’ approach?
\item \textbf{Technique}: what technique(s) will you use? Narration, hands-on, tinkering, exhibition? Analogic or digital? Will you use videos or projections?
\item \textbf{Material}: what material will you need? Projector, screen, speakers? Table, glue, paper? Copper tape, LEDs? What is the foreseen cost of the material? Is there special material that could be hard to find or is it all common material?
\end{itemize}

\noindent \textbf{Senses and notes}
\begin{itemize}
\item \textbf{Involved senses}: what senses will be involved? Sight and hearing? Touch, sigh and hearing? Smelling?
\item \textbf{Other notes}: are there other important things to consider? E.g., is a LIS (Italian Language of Sign) interpreter needed? Could there be criticalities? Are there problems to solve or doubts?

\end{itemize}

\subsection{Feedback Wheel}
\label{subsec:debrief}
The Feedback Wheel (Figure \ref{fig:temp}) is a collection of questions about users' emotions, feelings, perceptions, and criticism perceived during the activity:
\begin{itemize}
    \item How did you feel at the beginning? And then during the activity?
    \item What surprised you?
    \item What did you find difficult?
    \item Do you think other children might enjoy it?
    \item Any other suggestions?
\end{itemize}
 
\subsection{ICEBAGs}
\label{subsec:icebags}

ICEBAGS (Figure \ref{fig:temp}) is a short handbook for a good, accessible workshop. It lists best practices and best avoided for six activity areas:
\begin{itemize}

\item \textbf{Introduction}: How to start the workshop? How to introduce the workshop facilitator? How to introduce the rules, topics, and structure of the workshop? What to avoid?
\item \textbf{Conclusion:} How to conclude the workshop? How to check if the goals of the workshop are reached? How to improve the next workshop?
\item \textbf{Explanations \& Voice:} How to calibrate explanations to the audience? What to do if someone doesn’t know something? How to manage time? How to select the tone and volume of voice?
\item \textbf{Body language: }How to be welcoming with the body? What to avoid?
\item \textbf{Accessibility}: How to act in case of impaired users? How to make them feel comfortable about the workshop? How to equally offer the activity to all?
\item \textbf{Group dynamics}: What to do if someone looks uncomfortable doing the activity? What to do if we see a negative group dynamic? How to involve all users equally?
\item \textbf{Space arrangement}: How to arrange the space where the workshop takes place? How to place facilitator(s) and participants?
\end{itemize}

The shared materials include also the ICEBAGS produced in the 2024 edition.

\subsection{Questionnaires}
\label{quest}

\textbf{PhD candidates}
The questionnaires delivered to PhD candidates included the following questions:
\begin{enumerate}
    \item On a scale of 1 (= not at all) to 5 (= very much) and based on your previous knowledge and expectations, how useful was the course?
    \item On a scale of 1 (= not at all) to 5 (= very much) and based on your previous knowledge and expectations, how interesting was the course?
    \item Would you suggest other PhD candidates to attend this course in the future?
    \item What did you like about the course?
    \item What did you not like about the course?
    \item Are there other comments and suggestions that you would like to give us?
\end{enumerate}

\textbf{HS students}
The questionnaires delivered to HS students included the following questions:
\begin{enumerate}
    \item On a scale of 1 (= disappointing ) to 5 (= excellent), how was the experience in terms of content?
    \item On a scale of 1 (= disappointing ) to 5 (= excellent), how was the experience in terms of interactions with the trainers (researchers and PhD candidates)?
    \item On a scale of 1 (= disappointing ) to 5 (= excellent), how was the experience in terms of interactions with the other HS students?
    \item On a scale of 1 (= disappointing ) to 5 (= excellent), how was the experience in terms of interactions with the public?
    \item On a scale of 1 (= not at all) to 5 (= very much), how engaging did you find it to explore the Universe with all your senses?
    \item On a scale of 1 (= not at all) to 5 (= very much), how useful did you find it to explore the Universe with all your senses for understanding?
    \item After attending the Festival, how feasible do you think it is for blind or visually impaired individuals to study astronomy or work in the field of astronomy? [Possible answers: (1) I am now more confident that astronomy is accessible for blind and visually impaired individuals; (2) I am now less confident that astronomy is accessible for blind and visually impaired individuals; (3) My opinion has not changed]
    \item Do you have any other comments you would like to add?
\end{enumerate}

\textbf{Public}
The questionnaires delivered to the Festival participants included the following questions:
\begin{enumerate}
    \item Where are you from?
    \item How old are you?
    \item Are you a blind or visually impaired person? [yes/no]
    \item Is this your first time at the Festival? [yes/no]
    \item How did you hear about the Festival? [facebook / instagram / newspaper, TV, radio / from a friend or a relative / printed poster / advertising on public transport ]
    \item On a scale of 1 (= not at all) to 5 (= very much), what is your overall level of satisfaction with the Festival?
    \item On a scale of 1 (= not at all) to 5 (= very much), how engaging did you find it to explore the Universe with all your senses?
    \item On a scale of 1 (= not at all) to 5 (= very much), how useful did you find it to explore the Universe with all your senses for understanding?
    \item After attending the Festival, how feasible do you think it is for blind or visually impaired individuals to study astronomy or work in the field of astronomy? [Possible answers: (1) I am now more confident that astronomy is accessible for blind and visually impaired individuals; (2) I am now less confident that astronomy is accessible for blind and visually impaired individuals; (3) My opinion has not changed]
    \item On a scale of 1 (= not at all) to 5 (= very much), what is your level of satisfaction with the different events? [workshops / conferences / sky observations / exhibitions / shows / hike]
    \item Do you have any other comments you would like to add?
\end{enumerate}

\textbf{Teachers}
The questionnaires delivered to teachers participating in Phase 4 included the following questions:
\begin{enumerate}
    \item On a scale of 1 (= not at all) to 5 (= very much), how useful did you find the training course?
    \item On a scale of 1 (= not at all) to 5 (= very much), how interesting did you find the training course?
    \item How did you feel during the course (both during the training in Verona and the classroom training)?
    \item Would you recommend the course to other colleagues? [yes/no]
    \item What did you like?
    \item What did you not like?
    \item On a scale of 1 (= not at all) to 5 (= very much), did you find the hands-on activities useful for imparting knowledge to students?"
    \item On a scale of 1 (= not at all) to 5 (= very much), did you find the hands-on activities interesting for the students?
    \item Did you notice any differences in student learning during or after the hands-on activities compared to the lectures? If so, what were they?
    \item Did you notice any differences in student attitude during or after the hands-on activities compared to the lectures? If so, what were they?
    \item Have you used the hands-on activities with other classes outside of the training course?
    \item Do you plan to repeat the hands-on activities with other classes in the coming years?
    \item On a scale of 1 (= poor) to 5 (= excellent), how would you rate the training course?
    \item Do you have any other comments you would like to add?
\end{enumerate}

\subsection{Word clouds}
The words that were mentioned in the word clouds of Figure \ref{fig:word_clouds} are reported in Table \ref{tab:word_clouds}, divided in semantic group.
\begin{table}
    \centering
        \caption{Words used by high-school students in the word clouds. We divided them in five semantic categories to ease the analysis.}
    \begin{tabular}{|c|c|l|l|l|} \hline 
         \textbf{Knowledge}& \textbf{Relationships}& \textbf{Emotions/feelings}& \textbf{Self-growth}&\textbf{Other}\\ \hline 
         Convey&Collaboration&Amazement& Adaptation&Easy\\ \hline 
 Culture& Communication& Calm& Awareness&Marks\\ \hline 
 Delving& Diversity& Commitment& Enrichment&Star Wars\\ \hline 
 Discover& Entertaining& Curiosity& Experience&Synchrotron\\ \hline 
 Expertise& Express yourself& Empathy& Experimenting&Uniqueness\\ \hline 
 Interest& Friendship& Engagement& Improvisation&\\ \hline 
 Knowledge& Group& Extrovert& Maturity&\\ \hline 
 Learning& Interactions& Fatigue& Novelty&\\ \hline 
 Teaching& Meeting People& Fulfillment& Problem solving&\\ \hline 
         Understand&  Meetings& Fun& Responsibility&\\ \hline 
         Useful&  Opening& Initiative& &\\ \hline 
         &  Relationships& Inspiration& &\\ \hline 
         &  Sharing& Joy& &\\ \hline 
         &  Sociality& Passion& &\\ \hline 
         &  Teamwork& Patience& &\\ \hline 
         &  Welcoming& Satisfaction& &\\ \hline 
 & & Surviving& &\\ \hline 
         &  & Wonder& &\\ \hline
    \end{tabular}
    \label{tab:word_clouds}
\end{table}

\pagestyle{empty}

\end{document}